\documentclass[graybox]{svmult} % modified to show contents in one page
\usepackage{mathptmx}         % selects Times Roman as basic font 
\usepackage{helvet}           % selects Helvetica as sans-serif font
\usepackage{courier}          % selects Courier as typewriter font
\usepackage{makeidx}          % allows index generation
\usepackage{multicol}         % used for the two-column index
\usepackage[bottom]{footmisc} % places footnotes at page bottom
\makeindex             % used for the subject index
% ...
% ...
% ...
% ...
\smartqed  % flush right qed marks, e.g. at end of proof
\usepackage{epsfig}
\usepackage{amsmath,amssymb,amsbsy,natbib,extarrows,hhline,upgreek}

\begin{document}

%\title{Femtoscopy: Progress, Status and Outlook
%\thanks{Support was provided by the U.S. Department of Energy, Grant No. DE-FG02-03ER41259, and by the U.S. National Science Foundation, Grant No. PHY-0653432.}}
%\subtitle{on the eve of the LHC program}

\title*{Femtoscopically Probing the Freeze-out Configuration in Heavy Ion Collisions
\thanks{Support was provided by the U.S. Department of Energy, Grant 
No. DE-FG02-03ER41259, and by the U.S. National Science Foundation, Grant No. PHY-0653432.}}

\titlerunning{Femtoscopically Probing the Freeze-out Configuration}

\author{Michael Annan Lisa and Scott Pratt}
\authorrunning{M.A. Lisa and S. Pratt}

\institute{
Michael Annan Lisa \at
Department of Physics, Ohio State University \\
191 West Woodruff Ave \\
Columbus, OH 43210 \\
\email{lisa@mps.ohio-state.edu}
\and
Scott Pratt \at
Department of Physics and Astronomy, Michigan State University,\\
East Lansing, Michigan 48824 \\
\email{prattsc@msu.edu}           
}

\date{Received: date / Accepted: date}

\maketitle

\abstract{
Two-particle femtoscopy reveals the space-time substructure of the freeze-out configuration from heavy ion collisions.
Detailed fingerprints of bulk collectivity are evident in space-momentum correlations, which have been systematically
measured as a function of particle type, three-momentum, and collision conditions.  A clear scenario, dominated by
hydrodynamic-type flow emerges.  Reproducing the strength and features of the femtoscopic signals in models involves
important physical quantities like the Equation of State, as well as less fundamental technical details.
An interesting approximate ``factorization'' in the measured systematics suggests that the overall physical freeze-out
scale is set by final state chemistry, but the kinematic substructure is largely universal.  Referring to previous
results from hadron and lepton collisions, we point to the importance of determining whether these ``universal'' trends
persist from the largest to the smallest systems.  We review theoretical expectations for heavy ion femtoscopy at the
LHC, and point to directions needing further theory and experimental work at RHIC and the LHC.
}
%\keywords{HBT \and femtoscopy \and heavy ion collisions \and freeze-out \and flow}
%\PACS{25.75.Gz}

\setcounter{tocdepth}{4}
\clearpage
\tableofcontents
\newpage

\section{Introduction}
\label{sec:intro}

Amongst all the classes of observables analyzed in subatomic collisions, femtoscopy is 
unique as it directly addresses the space-time structure of the evolving system at the Fermi scale.
It exploits the fact that the likelihood to emit particle $a$ with momentum $p_a$ is modified by
the emission of particle $b$ with momentum $p_b$, so long as there is any structure-- due to 
quantum (anti-)symmetrization and/or final state interactions--
in the two-particle relative wavefunction $\phi\left(q,r\right)$, where $q=(p_a-p_b)/2$ and $r$ is
the relative space-time separation.  The $q$-dependence of the modification, which can be
measured, reveals the separation distribution of the particles.

The techqnique of measuring the arrival-time correlation of radio-wavelength photons to extract angular sizes of stars
was invented more than 50 years ago by Hanbury Brown and Twiss~\cite{HBT1,HBT2}.
Although the entire approach was initially greeted skeptically\footnote{Towards the end of his life, Hanbury Brown gave a retrospective
presentation~\cite{CRIS98} on this seminal work.  In its entirety, his listed abstract read 
``The talk will give a brief history of the early development of Intensity Interferometry and its subsequent battle against common sense.''
See also~\cite{Boffin}.},
the technique was used for years in the ground-based astronomy community.  

Five years later,
correlations of identical pions were shown to be sensitive
to source dimensions in proton-antiproton collisions by Goldhaber, Goldhaber,
Lee, and Pais (GGLP)~\cite{Goldhaber:1960sf}. 
In the 1970s, these methods were refined
by Kopylov and Podgoretsky~\cite{Kopylov:1972qw,Kopylov:1974uc}, Koonin,~\cite{Koonin:1977fh},
and Gyulassy~\cite{Gyulassy:1979yi}, and other classes of correlations were
shown to be useful for source-size measurements, such as strong and Coulomb
interactions.

While the ``GGLP effect'' in particle physics and the original ``HBT effect'' are in principle based on different theoretical frameworks~\cite{Baym:1997ce,Kopylov:1976},
in practice the extracted information and techniques are quite similar.  Thus, when in the 1980's experiments at the Bevalac~\cite{Fung:1978eq,Zajc:1984vb}
made clear that identical pion correlations yielded quantitative space-time information in nuclear collisions, the heavy ion community referred to such measurements
as ``HBT analyses.''  
Around the same time, the technique began to be applied to proton-proton and fragment-fragment correlations in low energy nuclear collisions~\cite{Boal:1990yh}.

HBT measurements were among the first results reported when
studies of truly relativistic heavy ion collisions began twenty years ago with the availability of Oxygen and Silicon beams SPS/CERN and AGS/BNL, respectively.
Since then, extracting space-time scales from two-particle correlations has become an integral part of the heavy ion program worldwide.  The availability
of high-statistics data from large-acceptance detectors and an continuing development of theoretical formalism has tremendously expanded the range of geometrical
information extracted from the correlations.  Moving far beyond the original ``source size'' measurements, modern analyses routinely extract the extent, shape,
orientation, and dynamical timescale of the emission zone.  This information has been systematically extracted and compared as a function of collision energy,
centrality and ion species.  It is further binned in terms of $p_T$, $y$ and $\phi$, and for a growing variety of particle species. 

Data systematics and model comparisons to RHIC, SPS, and AGS data have recently been reviewed in detail~\cite{Lisa:2005dd}.  Referring the reader there
for full details, we wish here to present a minimally self-contained overview of methodology and status of femtoscopy in heavy ion collisions on the eve of the LHC.
We will discuss generic features of the source in these collisions, which we consider firmly established.
Finally, we will briefly discuss some expectations for femtoscopy at the LHC, and open questions to be addressed there.

\subsection{The urgency of space-time in heavy ion studies}

The goal of the relativistic heavy ion physics program is the creation and study of color-deconfined bulk matter.
The hope is to to better understand and test QCD in the sector in which it is 1) most interesting and at the same time 2) most difficult to calculate.  It is most interesting in the low momentum transfer (alternatively large length scale or ``soft'') sector, since it is here where its most characteristic and important feature-- color confinement-- is manifest.  On the other hand, it is most theoretically difficult since the coupling between the colored objects in the theory is strong, rendering perturbative techniques ineffective. This feature has a major upside, however.  In particular, it gives hope that heavy ion collisions might produce a bulk system, at least partially thermalized, which can be described in the language of bulk quantities like temperature and pressure.
%; it makes little sense to discuss the ``matter'' created in collisions if this condition is not met.
%(color glass people would be offended by this remark)

%1) The defining feature of the Strong interaction is color confinement, a feature fundamentally characterized by a physical length scale.  Distinct from all other interactions, either fundamental or in effective field theories, this scale is not determined by the presumably Higgs-generated mass of an exchange boson, but by a unique feature of the theory-- the color charge carried by the exchange bosons themselves.  Any body of information brought to bear on QCD then, must include information directly related to physical length scales.  

The most celebrated early result from the RHIC program was the success of hydrodynamic calculations to describe the transverse momentum and azimuthal structure of measured hadron spectra.  While the focus has since shifted to
more refined and important details related to initial conditions, pre-equilibrium collectivity, finite viscosity and equations of state, there remains near-universal consensus that the measured soft-sector momentum distributions are clear evidence of bulk collective behaviour-- hydrodynamic flow.  However, projection of the seven-dimensional (space-time and momentum) emission pattern onto momentum-only axes represents only a minimal test of the complex, non-trivial geometric substructure predicted by hydrodynamics. In particular, collective flow is much more precisely characterized by signature correlations between momentum and coordinate space.
Femtoscopy provides the most direct experimental access to such patterns.  A solid case for a bulk system in heavy ion collisions depends crucially on such measurements.

Intrinsic thermodynamic quantities which characterize a state of matter, require knowing the volume of the sample of matter under study. The relationship between these quantities constitutes the Equation of State (EoS). Although much of the information we seek is for the static, equilibrium case, experimental information on QCD in this sector can only be obtained through the study of highly dynamic systems-- i.e. collisions.  Furthermore, the dynamics reflect the equation of state and viscosity. Without the spatial information, inferences of bulk quantities like pressure, energy density or entropy density would require untested assumptions about the volume, and without the temporal information, inferences into dynamics would be largely faith-based. 

The high inherent dimensionality to correlation functions (six dimensions for each species pair and for each beam energy, beam/target combination and centrality) make correlations the ultimate test for any model that claims to describe the evolution of a heavy ion collision. Thus, it is no surprise that femtoscopic data have proven the most challenging to explain of all heavy-ion observables. For these reasons femtoscopy is perhaps the most important soft-sector observable in relativistic heavy ion physics.

\subsection{The neccessity of systematics in heavy ion studies}

The physics program at the Relativistic Heavy Ion Collider (RHIC) at Brookhaven National Laboratory is
remarkably rich, thanks to the machine's unique ability to collide nuclei from $^1{\rm H}$ to $^{197}{\rm Au}$,
in fully symmetric (e.g. A+A or p+p) to strongly asymmetric (e.g. $d+Au$) entrance channels, over an energy
range spanning more than an order of magnitude.  
%%% the ``more than two orders of magnitude'' comes from Declan pointing out that in 2008, RHIC demonstrated ability to go down to sqrt(s)=5 GeV
The capability to collide polarized protons provides access to an entirely
new set of fundamental physics, not discussed further here.

Achieving the primary aim of
RHIC-- the creation and characterization of a color-deconfined state of matter and its transition back
to the confined (hadronic) state-- requires the full capabilities of RHIC.  In particular, comparisons of
particle distributions at high transverse momentum ($p_T$) from A+A and p+p collisions, probe the color-opaque
nature of the hot system formed in the collisions~\cite{Gyulassy:1990ye,Baier:2000mf,Adcox:2001jp}.
Comparison with reference $d+A$ collisions were
necessary to identify the role of initial-state effects in the spectra~\cite{Adams:2003im}.  Comparing anisotropic
collective motion from non-central collisions of different-mass initial states (e.g. Au+Au versus $Cu+Cu$)~\cite{Bhalerao:2005mm}
tests the validity of transport calculations crucial to claims of the creation of a ``perfect liquid'' at RHIC~\cite{Gyulassy:2004zy}.
Indeed, a main component of the future heavy ion program at RHIC involves a detailed energy scan, designed to
identify a predicted critical point in the Equation of State of QCD~\cite{Ritter:2006zz}.

The need for such systematic comparisons is not unique to RHIC, but has been a generic feature of all
heavy ion programs~\cite{Nagamiya:1988ge,Tannenbaum:2006ch}, from low-energy facilities like the NSCL (Michigan State),
to progressively higher-energy facilities at
SIS (GSI), the Bevatron/Bevalac (Berkeley Lab), AGS (Brookhaven), and SPS (CERN).  The nature of heavy ion physics
is such that little is learned through study of a single system.

The femtoscopic excitation function, in particular, was expected to exhibit an unambiguous signal of a hoped-for sharp phase transition, 
based on general grounds and essentially independent of a model~\cite{Rischke:1996em,Harris:1996zx,Bass:1998vz}.
The absence of such a signal was originally termed the ``RHIC HBT puzzle''~\cite{Heinz:2002un}.  However, our understanding of the
deconfinement transition at RHIC has evolved away from sharp first-order phase transition, thanks in part to the femtoscopic measurement.
In addition, more detailed refinements in the comparison of models with measured correlation functions, discussed below, have largely ameliorated this puzzle.
Nevertheless, we will mention further possible ``puzzles'' in direct comparisons of femtoscopy from heavy ion versus elementary particle collisions.

\subsection{Organization of this paper}

In Section~\ref{sec:primer}, we present a brief primer on two-particle femtoscopy, sketching the theoretical formalism and connection of the correlation
function to the structure of the source function.   %%%% this should include discussion of homogeneity regions and imaging too
In Section~\ref{sec:flow}, we review the systematics mapped out by a broad range of femtoscopic measurements, and the scenario which
clearly emerges.
There, we also briefly discuss what
has been observed in high energy hadron-hadron and lepton-induced collisions.  In Section~\ref{sec:bulk},
we discuss conclusions drawn from parametric and transport model comparisons so far.
In Section~\ref{sec:Expectations}, we review theoretical expectations for
heavy ion collisions at the LHC, expected in just a few years.
In Section~\ref{sec:summary}, we summarize.

\section{A Femtoscopy Primer}
\label{sec:primer}

Here, we give a very brief overview of the formalism of femtoscopic correlation functions.
For a more complete discussion of the theoretical and experimental details, see~\cite{Lisa:2005dd}.

\subsection{What is measured - correlation functions and source functions}

Femtoscopic analyses are based on two-particle correlation measurements at small relative momentum. For particles of types $a$ and $b$, the
six-dimensional correlation function is typically expressed in terms of the relative and total momentum,
\begin{eqnarray}
C^{ab}({\bf P},{\bf q}) &=&\frac{dN^{ab}/(d^3p_ad^3p_b)}
{(dN^a/d^3p_a)(dN^b/d^3p_b)}\, ,
\label{eq:cdef}
\\
\nonumber
P&\equiv& p_a+p_b,~~
q^\mu=(p_a-p_b)^\mu/2-\frac{(p_a-p_b)\cdot P}{2P^2}P^\mu\, .
\end{eqnarray}
The last term in the definition of the relative momentum $q$ projects out the temporal component in the two-particle rest frame. At times, we will use $Q\equiv 2q$ to refer to the relative momentum.

Experimentally, the formal definition of the correlation function in Equation~\ref{eq:cdef} is seldom used in heavy ion physics.
Instead, it is approximated by
\begin{equation}
\label{eq:ExperimentalDefinition}
C^{ab}_{\bf P}({\bf q}) = \frac{A^{ab}_{\bf P}({\bf q})}{B^{ab}_{\bf P}({\bf q})}\cdot \xi_{\bf P}({\bf q}),
\end{equation}
where ${A^{ab}_{\bf P}({\bf q})}$ is the signal distribution, ${B^{ab}_{\bf P}({\bf q})}$ is the reference or background distribution which is ideally
similar to $A$ in all respects except for the presence of femtoscopic
correlations.  In heavy ion collisions, $B$ is usually formed via the event-mixing technique~\cite{Kopylov:1974th}, in which ${\bf q}$ (or any pair-wise
variable) is formed when particles $a$ and $b$ come from different events; in contrast, the histogram ${A^{ab}_{\bf P}({\bf q})}$ represents the distribution
when the particles come from the same event.
Except in measurements in which the acceptance is extremely limited~\cite{Zajc:1984vb} or for very low multiplicities~\cite{Rebreyand:1990xx}, 
Equations~\ref{eq:cdef} and~\ref{eq:ExperimentalDefinition} yield practically identical results~\cite{Lisa:1991zz}.
In any event, the reference distribution $B$ should account for
effects of detector acceptance and efficiency, and fluctuaions in reaction-plane orientation, multiplicity, and vertex position;
accounting for such effects are most naturally done with the event-mixing technique.
$\xi_{\bf P}({\bf q})$ represents a potential correction factor introduced to
compensate for non-femtoscopic correlations present in the signal that are not
fully accounted for in the background as well as artifacts resulting, e.g., from
finite resolution and contamination.
See~\cite{Lisa:2005dd} and references therein, for a more detailed discussion.

The designation of the total momentum ${\bf P}$ by a subscript in Equation~\ref{eq:ExperimentalDefinition} reflects
the fact that usually experimental correlation functions are formed in terms of ${\bf q}$ and analyzed separately
for various selections of the total momentum ${\bf P}$.
As discussed in Section~\ref{sec:intro}, femtoscopic correlations arise only through the spatially-dependent
two-particle relative wavefunction $\phi_{a,b}({\bf q},{\bf r})$, which vanishes as $|{\bf q}\rightarrow\infty|$.
Thus, for large relative momentum, the correlation function approaches a constant value; typically, $C(|{\bf q}\rightarrow\infty|)$
is normalized arbitrarily to unity.  The presence of other correlations (e.g. due to momentum conservation~\cite{Chajecki:2008vg})
can complicate matters, especially for small systems.

Femtoscopic correlations are expected to be described by the Koonin equation \cite{Koonin:1977fh},
\begin{eqnarray}
\label{eq:koonin}
C^{ab}({\bf P},{\bf q})&=&\int d^3r'~S^{ab}({\bf P},{\bf r}')\left|\phi({\bf q}',{\bf r}')\right|^2,\\
\nonumber
S^{ab}({\bf P},{\bf r}')&\equiv&
\frac{\int d^4x_a d^4x_b
s_a\left(\frac{m_a}{m_a+m_b}{\bf P},x_a\right) s_b\left(\frac{m_b}{m_a+m_b}{\bf P},x_b\right)
\delta^3\left({\bf r}'-[{\bf x}'_a-{\bf x}'_b]\right)}{\int d^4x_a d^4x_b
s_a\left(\frac{m_a}{m_a+m_b}{\bf P},x_a\right) s_b\left(\frac{m_b}{m_a+m_b}{\bf P},x_b\right)}.
\end{eqnarray}
Here, the squared relative outgoing two-particle wave function serves as a weight which enhances or reduces the probability for emitting two particles depending on their separation. In Eq. (\ref{eq:koonin}) ${\bf r}'$ refers to the separation of the two particles in the two-particle rest frame and ${\bf q}'$ is the relative momentum in that frame. The probability of emitting a single particle of momentum $p$ from space-time point $x$ is $s(p,x)$, and is evaluated at a momentum such that each particle has the same velocity and such that the total momentum is ${\bf P}$. Justification of the Koonin equation requires assuming that the sources are highly uncorrelated (chaotic), that the phase space density is not too large,  and that the relative momentum is small. A detailed discussion is provided in \cite{Lisa:2005dd}. 

Since the separation of two particles of the same velocity is independent of time, the source function can also be written in terms of the asymptotic phase space densities in the pair frame (where both momenta are zero),
\begin{eqnarray}
\label{eq:kooninb}
S^{ab}({\bf P},{\bf r}')&\equiv&
\frac{\int d^3x'_a d^3x'_b~
f_a\left({\bf p}'_a=0,{\bf x}'_a,t\right) f_b\left({\bf p}'_b=0,{\bf x}'_b,t\right)
\delta^3\left({\bf r}'-[{\bf x}'_a-{\bf x}'_b]\right)}
{\int d^3x'_a d^3x'_b~
f_a\left({\bf p}'_a=0,{\bf x}'_a,t\right) f_b\left({\bf p}'_b=0,{\bf x}'_b,t\right)},
\end{eqnarray}
assuming $t$ is any time after all particles have been emitted. It is more accurate to state that femtoscopy measures the structure of outgoing phase space clouds of a given velocity, rather than claiming that the source is measured. 

Since correlations are six-dimensional objects, there is no hope to extract all seven 
dimensions of information in $s({\bf p},x)$. The separation ${\bf x}'=\gamma({\bf x}-{\bf v}_{\bf P}t)$, the spatial separation along the direction of ${\bf P}$,  is inextricably linked to the time. Nonetheless, if one assumes that the spatial separation along ${\bf P}$ is distributed similarly to one of the transverse
separations, one can infer the temporal information by considering the difference. Such assumptions usually become trustworthy only for long lifetimes, as has been measured femtoscopically in the case for heavy ion collisions at low energy~\cite{Lisa:1993xx,Lisa:1994zz,Kotte:1997kz}, $\lesssim 50A$ MeV.  We discuss this further in Section~\ref{sec:flow}, in relation to the ratio $R_{out}/R_{side}$.

Furthermore, since one can only infer information about the relative separation of two particles with a given velocity, one measures a subset of the entire source, even if the correlation measurements span all of momentum space. The subset of the volume relevant for femtoscopy is referred to as the {\it region of homogeneity}~\cite{Akkelin:1995gh,Lisa:2005dd}, and can strongly vary with ${\bf P}$, especially in the presence of collective flow, as discussed in Section~\ref{sec:flow}.

From viewing Equations~\ref{eq:koonin} and~\ref{eq:kooninb} it is clear that correlations can, at best, provide the {\it source function} $S({\bf P},{\bf r})$, and the principal role of femtoscopy is to discern all possible information about the source function from the correlation function. 
%%%Since the total momentum appears on both sides of Equation~\ref{eq:koonin}, the principal challenge for most analyses is the invert 
%%%the convolution relating the relative momentum and the relative position.

Rather than expressing the correlation function in terms of the three components of ${\bf q}$ or the source function in terms of the three components of ${\bf r}$, one can decompose the angular information using a basis of either Cartesian or spherical harmonics~\cite{Danielewicz:2006hi,Chajecki:2005qm,Chajecki:2008vg}. In Cartesian harmonics the components are labeled by $\vec\ell\equiv(\ell_x,\ell_y,\ell_z)$, with $\ell=\ell_x+\ell_y+\ell_z$, and the decomposition of an angular function $F(\Omega)$ is defined by
\begin{eqnarray}
F_{\vec{\ell}}&=&\frac{(2\ell+1)!!}{\ell!}\int \frac{d\Omega}{4\pi} F(\Omega) A_{\vec\ell}(\Omega),\\
\nonumber
F(\Omega)&=&\sum_{\vec\ell} F_{\vec\ell}\hat{\Omega}_x^{\ell_x}\hat{\Omega}_y^{\ell_y}\hat{\Omega}_z^{\ell_z},
\end{eqnarray}
where $\hat{\Omega}$ is a unit vector. Some examples of the Cartesian harmonics, $A_{\vec\ell}(\Omega)$, are: $A_{0,0,0}=1, A_{1,0,0}(\hat{\Omega})=\hat{\Omega}_x, A_{2,0,0}(\hat{\Omega})=\hat{\Omega}_x^2-(1/3), A_{1,1,0}(\hat{\Omega})=\hat{\Omega}_x\hat{\Omega}_y, \cdots$. The convenience of such decompositions derives from the property that a given angular component of the correlation function has a one-to-one link to the corresponding element of the source function,
\begin{eqnarray}
\label{eq:imageCS}
C_{\vec\ell}(q)&=&\int 4\pi r^2dr~K_\ell(q,r)S_{\vec\ell}(r),\\
\nonumber
K_{\ell}(q,r)&\equiv&\frac{1}{2}\int d\cos\theta_{qr}P_{\ell}(\cos\theta_{qr}) |\phi(q,r,\theta_{qr})|^2,
\end{eqnarray}
where $\theta_{qr}$ is the angle between ${\bf q}$ and ${\bf r}$. Equation (\ref{eq:imageCS}) makes it possible to evaluate the correlation coefficients $C_{\vec\ell}$ component-by-component. Decompositions are also an efficient way to view features of the correlation arising from a specific symmetry. For instance, in a correlation function of non-identical particles, it is insightful to analyze the $(\ell_{\rm out}=1,\ell_{\rm side}=\ell_{\rm long}=0)$ projections. 
Spherical harmonics have the same convenient properties, and several simplifying symmetries~\cite{Chajecki:2008vg}

Any transport theory can generate emission probabilities, and can therefore provide a source function from which to calculate correlation functions; details of the calculation are discussed in~\cite{Lisa:2005dd}.  Comparisons of the calculated correlation functions to the measured ones represent the best possible test of the space-time structure of the model's freezeout configuration. However, it is often useful (for example when the calculated and measured correlation functions disagree) to parameterize source function to identify its dominant features, such as resonance tails, long lifetimes, or anisotropic shapes.  The extracted parameters, which we discuss next, are typically what are published and compared between experiments or to theory.

\subsection{Representation of the source function}

Here, we briefly review parameters and formalism commonly used to present and summarize geometric information from femtoscopic measurements.

\subsubsection{Reference frames and coordinate systems}
\label{sec:OutSideLongEtc}

The most common coordinate system in which to decompose the relative momentum is the Bertsch-Pratt
system~\cite{Bertsch:1988db,Pratt:1990zq}, in which ${\bf q}=\left(q_{out},q_{side},q_{long}\right)$.  The ``long'' axis is oriented along the beam direction; the transverse momentum vector ${\bf P_T}$ (i.e. the projection of the total momentum perpendicular to the beam) defines the ``out'' direction; the ``side'' direction is perpendicular to these. These are indicated in Figure~\ref{fig:blastcartoon}.

The correlation function is generally presented in one of three reference frames.  The Longitudinally Co-Moving System (LCMS) is defined such that the pair's total momentum along the beam direction vanishes (${\bf P_z}=0$).  This is commonly used at ultra-relativistic energies, for which the
system is approximately boost-invariant.  In a boost-invariant scenario, pairs with a given velocity in the beam direction $V_z$ are, on average, emitted from source elements moving with the same $V_z$. Since comparisons to $e^+-e^-$ collisions are increasingly interesting, as we discuss later, one should be aware that in these measurements, the LCMS system is defined such that the pair momentum along the event's thrust axis vanishes~\cite{Kittel:2001zw}.

In some analyses, an additional transverse boost is made to 
the Pair Center of Mass System (PCMS), in which the total momentum of the pair vanishes.
This differs from the LCMS only by a Lorentz factor $\gamma$ in the outward direction, but some prefer the PCMS since
the relative wave function is most conveniently expressed in this frame.  For instance, a sharp resonance peak
is no longer sharp if the correlation is viewed away from the pair frame.  The PCMS is usually used in analyses of
non-identical particle correlations (in which the relative momentum is sometimes denoted ${\bf k^*}$) and when
performing ``imaging'' fits as described below.

At low collision energies, for which boost invariance is a poor approximation, the CMS of the colliding system (the lab frame, for symmetric collisions at a collider)
is often used~\citep[e.g.][]{Lisa:2000no}; this would be the natural frame for a source with no longitudinal flow.

\subsubsection{Gaussian parameterizations}
\label{sec:GaussianParameterizations}

To extract the dominant scales and shape of the homogeneity region, a Gaussian source function is often used.  In general, the source function
of the separation ${\bf r'}\equiv{\bf x'_a}-{\bf x'_b}$ is
\begin{equation}
\label{eq:GeneralGaussian}
S\left(P,r'\right) \sim \exp\left[-\sum_{i=o,s,l}\sum_{j=o,s,l}\left(r'_i-\Delta_i\right)\left(r'_j-\Delta_j\right)\cdot\left[(R^2)^{-1}\right]_{i,j}/4\right] ,
\end{equation}
where $o,s,l$ stand for the ``out,'' ``side'' and ``long'' vector components.
Nine quantities characterize the distribution: the vector difference between
average emission points $\pmb{\Delta}$ and a symmetric $3\times 3$ covariance matrix $R^2$
\begin{align}
 \pmb{\Delta} & \equiv \langle {\bf x'_a}-{\bf x'_b}\rangle , \\
 R^2_{i,j} & \equiv \langle \left( r'_i-\Delta_i\right)\left( r'_j-\Delta_j\right) \rangle ,
\end{align}
where $\langle\dotsb\rangle$ indicates an average over the source function $S^{ab}$.  An implicit dependence of $\pmb{\Delta}$ and $R^2$ on ${\bf P}$
has been suppressed.
In most analyses, several of these nine factors vanish due to symmetry~\citep[e.g.][]{Heinz:2002au,Brown:2005ze}.

{\bf \underline{Identical particles}}

For identical particles ($a=b$), $\pmb{\Delta}={\bf 0}$ trivially, and the covariance matrix is typically
expressed in terms of the six ``HBT radii parameters'' $R^2_{i,j}$. For the commonly-measured two-pion correlations, they are usually extracted by fitting the measured correlation function with the functional form
\begin{equation}
C_{\bf P}\left({\bf q}\right) = N\left[\lambda K_{\rm Coul}\left({\bf Q}\right)\left(1+e^{-Q_iQ_jR^2_{ij}}\right)+\left(1-\lambda\right)\right] ,
\end{equation}
where $N$ is the overall normalization discussed above. Here, $K_{\rm coul}(q)$ is a factor, which approximately accounts for the Coulomb interaction between pions \cite{Bowler:1991vx,Sinyukov:1998fc}. For fully chaotic emission from a purely Gaussian source, and with perfect experimental particle identification (PID), the factor $\lambda$ should be unity. If experimental identification is perfect and the source is purely chaotic, $\lambda$ represents the fraction of correlated pairs, i.e., the fraction of pairs where neither pion is emitted from far outside the source being described by the Gaussian. For instance, if 25\% of the pions were either emitted from long-lived resonances or were mis-identified electrons, one would expect the lambda parameter to be $0.75^2$. Typically $\lambda\sim 0.3-0.7$, and in a typical fit to a measured correlation function, the covariance between the different ``HBT radii'' is small or zero, while the covariance between radii and $\lambda$ is usually slightly positive.

The diagonal components $R^2_o$, $R^2_s$ and $R^2_l$ characterize the lengths of the major axes of the homogeneity region of particle $a$, as shown in Figure~\ref{fig:blastcartoon}. The off-diagonal components $R^2_{os}$, $R^2_{sl}$ and $R^2_{ol}$ quantify the orientation of these axes relative to the ``out'', ``side'' and ``long'' directions; they encode important information about femtoscopic anistropies of the collision. The azimuthal dependence of $R^2_{os}$, relative to the reaction-plane, probes the transversely anisotropic shape of the source~\cite{Heiselberg:1998es,Lisa:2000ip};
it is the coordinate-space analog of the momentum anisotropy $v_2$. Finally, the azimuthal dependence of 
$R^2_{ol}$ $R^2_{sl}$ probe the spatial ``tilt'' of the collision zone~\cite{Lisa:2000ip}; it is the coordinate-space analog of directed flow $v_1$.

As a dynamic probe, the rapidity dependence of $R^2_{ol}$ reveals the extent to which the dynamics of the colliding system is boost-invariant~\cite{Tomasik:2002rx}. A finite value of this parameter, as measured in LCMS indicates a breakdown of boost invariance. For central collisions, an alternate Gaussian source parameterization is sometimes used, instead, to probe boost-invariance. Using the Yano-Koonin parameterization~\cite{Yano:1978gk,Wu:1996wk}, the correlation function is fit with the form~\cite{Heinz:1996qu}
\begin{equation}
C_{\bf P}\left({\bf Q}\right) = 
N\left[\lambda K_{\rm Coul}  \left({\bf q}\right)\left(1+e^{-R^2_\perp q^2_\perp - R^2_{||}(Q^2_{||}-Q^2_0)-(R^2_0-R^2_{||})(Q\cdot u)^2}\right)+\left(1-\lambda\right)\right] .
\end{equation}
Here, $u=\gamma (1,0,0,v_{YK})$, and $v_{YK}$ is the velocity of the source element, along the beam direction, which emitted the pions. In a boost-invariant system this velocity is identical to the longitudinal velocity of the pion pair.

{\bf \underline{Non-identical particles}}

Distinguishable particles ($a\neq b$) with the same velocity can have different average space-time freezeout positions, as quantified by the displacement vector $\pmb{\Delta}$ in Equation~\ref{eq:GeneralGaussian}.  For boost-invariant sources, $\Delta_{l}$ vanishes, as does $\Delta_{s}$ for analyses integrated over azimuth.  Few measurements of the displacements have been performed to date; only displacements along the pair velocity have so far been published at low~\cite{Cornell:1996xx,Kotte:1999gr,Gourio:2000tn,Ghetti:2001gm} and RHIC~\cite{Adams:2003qa} energies.

\subsubsection{Non-Gaussian parameterizations}
\label{sec:nonGauss}

While Gaussian scale and covariance parameters capture the bulk properties of the emitting source, there
is no reason to expect $S^{ab}\left({\bf r'}\right)$ to be purely Gaussian.  Indeed, discrepancies have long
been noted between measured correlation functions and fits using Equation~\ref{eq:GeneralGaussian}.  These
discrepancies are not large, but are to be expected.
For example, exponentially decaying resonance feed-down contributions will naturally lead to long-range tails
in the emission probability, creating a ``core-halo''~\cite{Csorgo:1994in} structure.  The ``halo'' affects
the low-$|{\bf q}|$ structure of the correlation function.  Until recently, such structure could not be resolved,
and the halo mostly affected the $\lambda$ factor in Gaussian fits~\cite{Csorgo:1994in}.  However, high-resolution
data at RHIC has allowed investigation into the source halo.

Various non-Gaussian functional forms and expansions have been proposed \cite{Csorgo:2003uv}, which fit the data better than a pure Gaussian, 
for example Edgeworth expansions \cite{Adams:2004yc}, or double-Gaussian fits. However recently there has been considerable activity in 
``imaging'' the source~\cite{Verde:2001md,Panitkin:2001qb,Brown:2000aj,Brown:1999ka,Brown:1997ku,Chung:2002vk}. 
As with the Gaussian fits discussed
above, imaging is essentially a multi-parameter fit of a calculated correlation function to the measured one.
The calculated correlation function is generated through Equation~\ref{eq:koonin}, where the source function $S^{ab}(r')$
is parameterized as a sum of spline functions.
The weights in this sum are the fit parameters varied to minimize $\chi^2$.
Considerable mathematical and numerical machinery has
been developed in order to achieve a stable fit-- including setting bin sizes and stationary points (or ``knots'') by hand--
and in practice the technique is nontrivial.  It may be fairly said that properly imaging a source remains something of an art
at this point. 
Mutli-dimensional imaging makes use of the harmonic decompositions of Equation~\ref{eq:imageCS}.  First experimental results
are discussed in Section~\ref{sec:flow}.

%In the last year, 
%three dimensional images have been created for the first time from data \cite{Afanasiev:2007kk} using angular decompositions of the source 
%function as described in Eq. (\ref{eq:imageCS}). Imaging and fitting to non-Gaussian forms have consistently revealed non-Gaussian halos 
%to the source function, which, are consistent with expectations from resonance decays and collective 
%motion~\cite{Adler:2006as,Afanasiev:2007kk,Brown:2007raa}.

\section{Present Status: The Scenario Supported by Femtoscopic Data}
\label{sec:flow}

In the absence of space-momentum correlations, $s({\bf p},x)$ would factorize into a function of ${\bf p}$ multiplied 
by a function of $x$, and all source parameters would be independent of the pair momentum ${\bf P}$ and independent of particle 
species. Collective hydrodynamic-like flow is a principal source of such correlations, but not the sole cause. For instance, 
jets represent a particularly dramatic example of space-time correlations as the particles with high transverse momentum 
hadronize further away in the outward direction. Resonant decays also provide correlations to some degree. Given that the 
characteristic mean free paths at early times are much shorter than the size of the fireball, the language of hydrodynamics 
is usually invoked to describe space-momentum correlations for low to mid-$p_t$ particles in heavy-ion collisions. 
Generically, the femtoscopic data paint a clear scenario dominated by hydrodynamic-like flow.  Such 
a source leaves  several unique femtoscopic fingerprints.  In this section,
we summarize these signatures, and see how each of them has been observed in the data so far.

\subsection{Longitudinal Flow: Rapidity and $m_T$ Systematics}
\label{sec:longitudinalFlow}

At ultrarelativistic energies, one expects a source extended-- and with strong flow-- along the beam axis;
this is in contrast to lower energies, where three-dimensional radial flow is observed~\cite{Bondorf:1978kz,Siemens:1978pb,Lisa:1994yr}.
Indeed, many models assume pure boost invariance, at least at midrapidity, in which the space-time rapidity equals the momentum rapidity.
%Boost invariance implicitly requires strong longitudinal flow.
%%%???  why do you say this, Scott?  A coordinate system doesn't assume dynamics ???and the LCMS coordinate system specifically 
%%%assumes boost invariance. 
Rapidity distributions alone suggest approximate boost invariance at midrapidity, but femtoscopic probes
provide stronger geometric support.
At RHIC and SPS, a boost-invariant scenario is supported by both
Yano-Koonin velocities~\cite{Kadija:1996wy,Appelshauser:1997rr,Antinori:2001yi,Back:2004ug} 
and the Bertsch-Pratt cross terms~\cite{Alt:2007uj} (c.f. Section~\ref{sec:GaussianParameterizations}) within a unit of midrapidity.  At more forward rapidities, a breakdown of
this scenario is revealed~\cite{Miskowiec:1996xa,Kadija:1996wy,Appelshauser:1997rr,Alt:2007uj} below RHIC energies.

\begin{figure}
\begin{minipage}[t]{0.52\textwidth}
%\centerline{\includegraphics[width=0.95\textwidth]{Figures/LisaFig08.eps}}
\epsfig{file=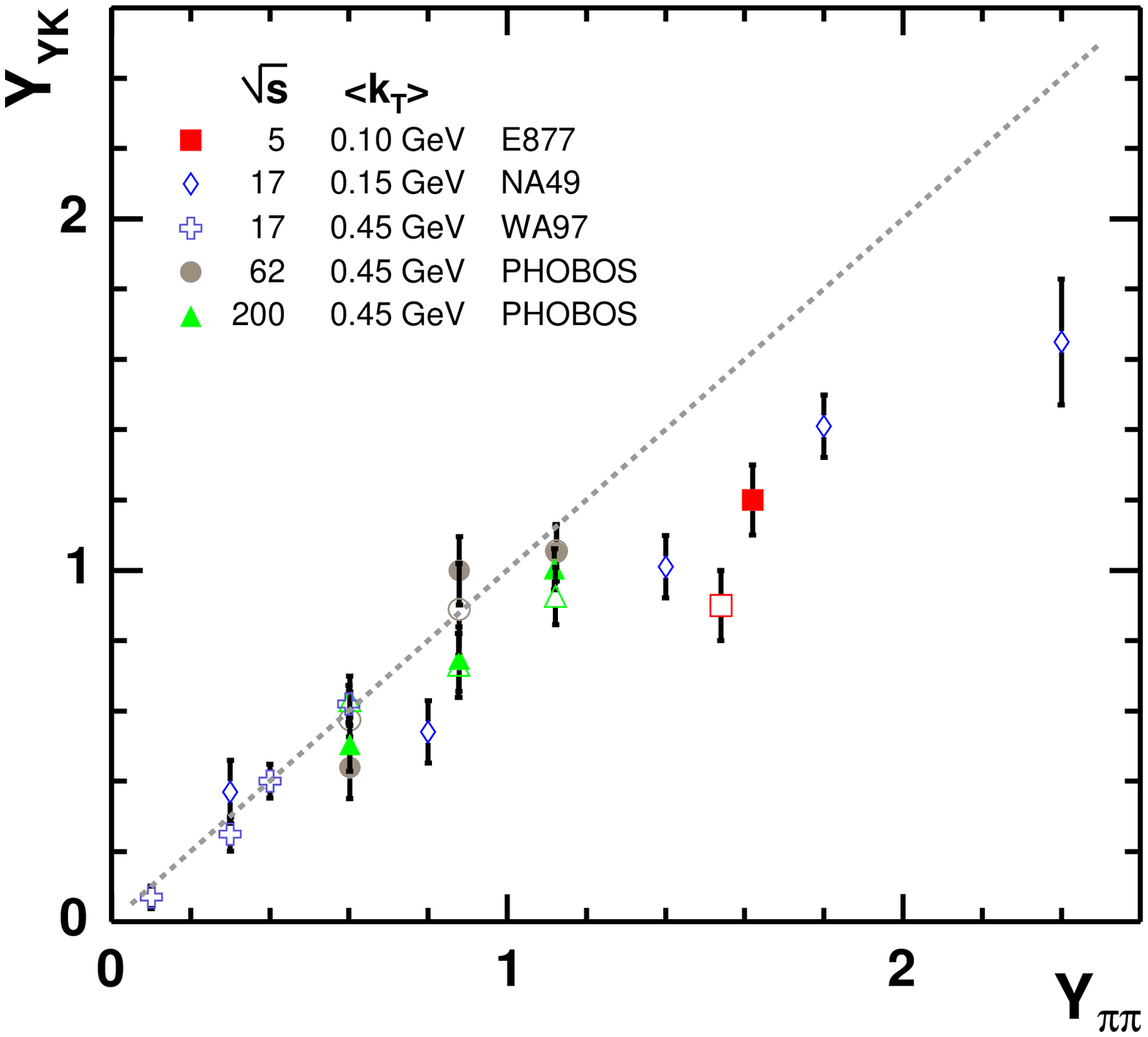,width=0.95\textwidth}
\caption{\label{fig:heavyIonYK}
Yano-Koonin rapidity versus pair rapidity, for pion pairs from central collisions, measured for heavy ion collisions
over a range of energies.  Open (closed) symbols show results for negative (positive) pion pairs.
(Figure from \cite{Lisa:2005dd})
}
\end{minipage}
\hspace{\fill}
\begin{minipage}[t]{0.44\textwidth}
\centerline{\includegraphics[width=0.95\textwidth]{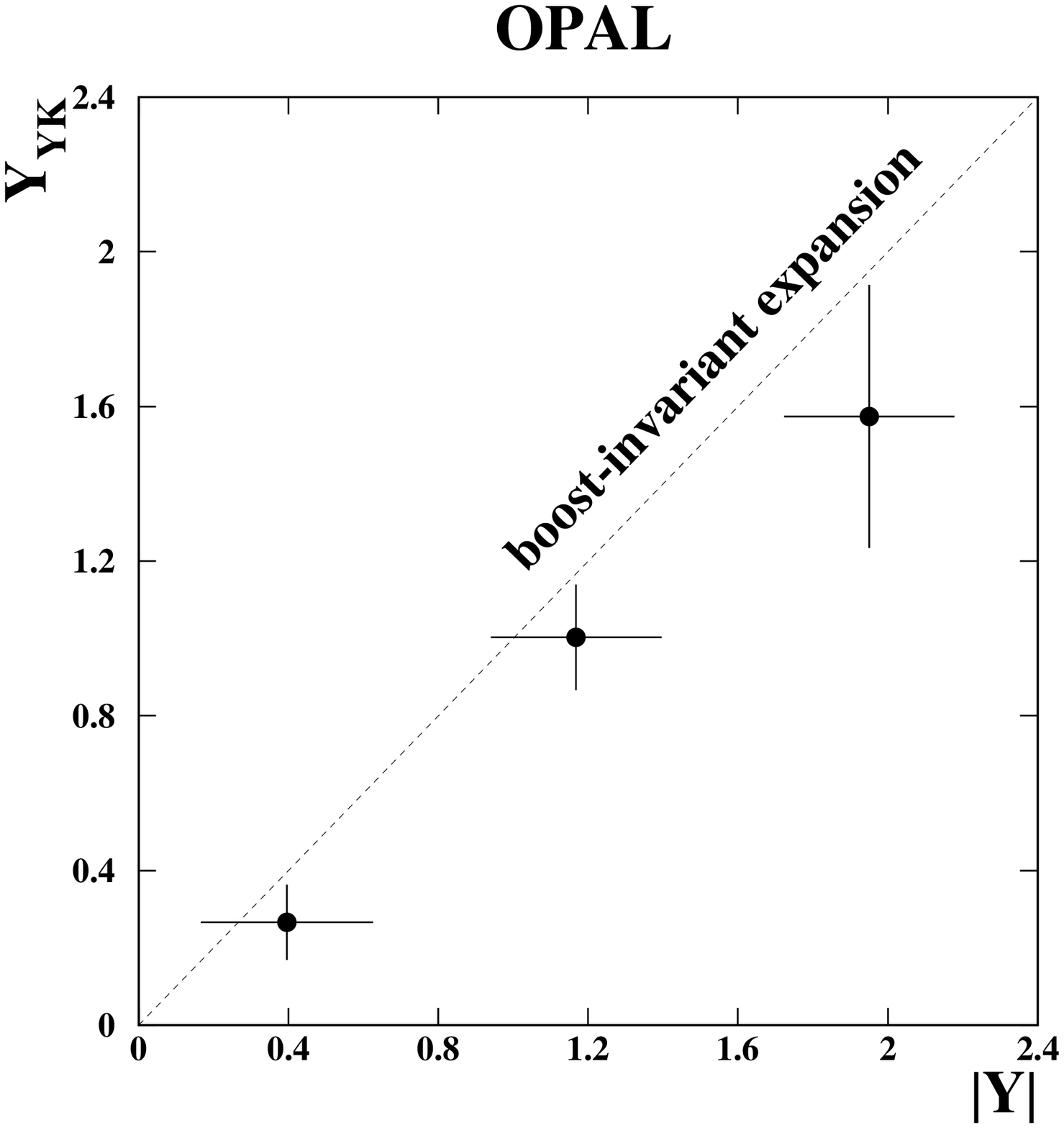}}
\caption{\label{fig:OPALYK}
Yano-Koonin rapidity versus pair rapidity, for pion pairs from $e^+-e^-$ collisions at LEP, measured by
the OPAL collaboration.  From~\cite{Abbiendi:2007he}.
}
\end{minipage}
\end{figure}

As seen in Figure~\ref{fig:heavyIonYK}, the Yano-Koonin source velocity (c.f. Section~\ref{sec:GaussianParameterizations}) is almost
identical to that of the pion pair, for a broad range of collision energies.
Figure~\ref{fig:OPALYK}
shows the same analysis in $e^+-e^-$ collisions by the OPAL Collaboration at LEP~\cite{Abbiendi:2007he}; here, the analysis is performed along the event
thrust direction, which is a more natural symmetry axis than the beam direction.

\begin{figure}
\centerline{\includegraphics[width=0.75\textwidth]{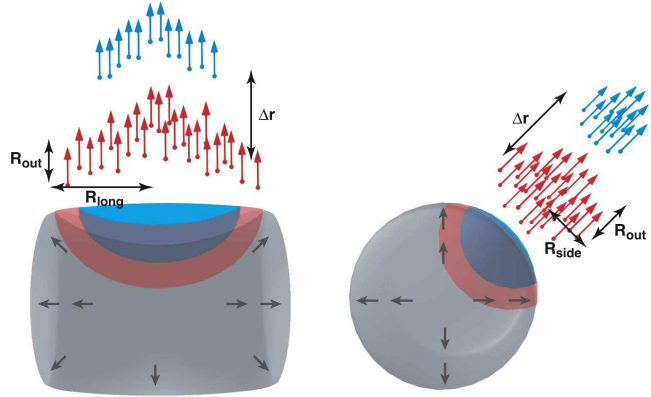}}
\caption{\label{fig:blastcartoon}
Emission zones for higher mass (blue) and lower mass (red) particles, as seen from alongside the beam (left) and looking down the beam pipe (right). Due to collective expansion particles of a given velocity come from a subset of the overall region, referred to as the region of homogeneity. Particles of higher mass, or higher transverse mass, have smaller thermal velocities and are thus confined to a smaller region by collective flow. For non-identical particles with velocities ahead of the surface, the more massive species are more confined to the surface, which leads to an offset $\Delta {\bf r}$ of the centroid of two outgoing phase space clouds. (Figure from \cite{Lisa:2005dd})
}
\end{figure}

Perhaps the best known, and the most misunderstood, signature of collective flow is known as $m_t$ scaling~\cite{Makhlin:1987gm}, 
which often simply refers to the fall of the Gaussian radius parameters with $p_t$. However, the source of this behavior is very 
different for $R_{\rm side}$ and $R_{\rm out}$ vs. $R_{\rm long}$. The term $m_t$ scaling refers only to the $p_t$ dependence of 
$R_{\rm long}$, and is inherently a relativistic effect. If one considers a purely longitudinal Bjorken expansion~\cite{Bjorken:1982qr} 
where the matter thermally dissociates at a fixed proper time $\tau=\sqrt{t^2-z^2}$, particles of transverse momentum $p_t$ and zero 
rapidity are emitted from position $z=\tau\sinh\eta$, where the collective rapidity is also $\eta$ with probability 
\begin{eqnarray}
\label{eq:slong}
\frac{dN}{dzd^3p}&\sim e^{-m_t\cosh\eta/T} \sim  e^{-m_t\sqrt{1+\left(z/\tau\right)^2}/T} .
\end{eqnarray}
For large $z$, the distribution falls off exponentially with a characteristic length, 
\begin{equation}
\label{eq:mT}
R_{\rm exp}=\frac{T\tau}{m_t},
\end{equation}
whereas for small $z$ the curvature mimics a Gaussian with a characteristic radius parameter~\cite{Makhlin:1987gm}
\begin{equation}
\label{eq:rootmT}
R_{\rm long}=\tau\sqrt{T/m_t}.
\end{equation}
This last relation motivated the term ``$m_t$ scaling'', as it suggests that $R_{\rm long}$ would scale as 
$1/\sqrt{m_t}$. 
The effect is illustrated in Fig. \ref{fig:blastcartoon}.

Indeed the $p_T$ dependence of $R_{long}$, seen in Figure~\ref{fig:BWfit}, goes as
$m_T^{-\alpha}$, where $\alpha$ is slightly
larger than 0.5~\cite{Lisa:2005dd}, as expected due to the additional effect of transverse flow~\cite{Herrmann:1994rr,Wiedemann:1995au,Retiere:2003kf}.
Lepton collisions at LEP display an $m_T$-dependence of the radii similar to that from heavy ion collisions, however; compare Figures~\ref{fig:BWfit}
and~\ref{fig:OPALkT}.

\begin{figure}
\begin{minipage}[t]{0.48\textwidth}
\centerline{\includegraphics[width=0.95\textwidth]{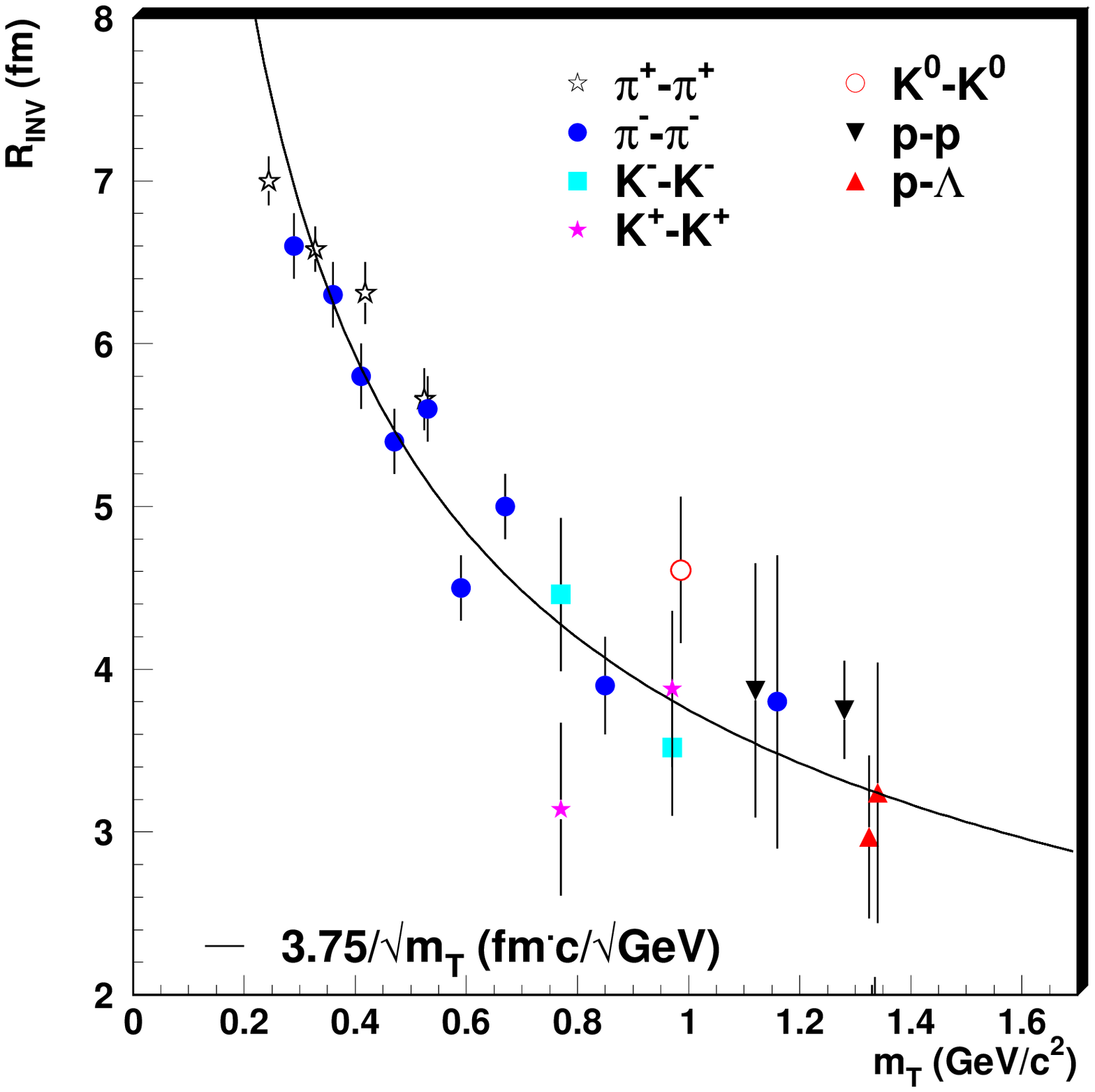}}
\caption{\label{fig:heavyIonMt}
One-dimensional Gaussian radii for source of various particles from central Au+Au collisions at RHIC.
(Figure from \cite{Lisa:2005dd}.)
}
\end{minipage}
\hspace{\fill}
\begin{minipage}[t]{0.48\textwidth}
\centerline{\includegraphics[width=0.95\textwidth]{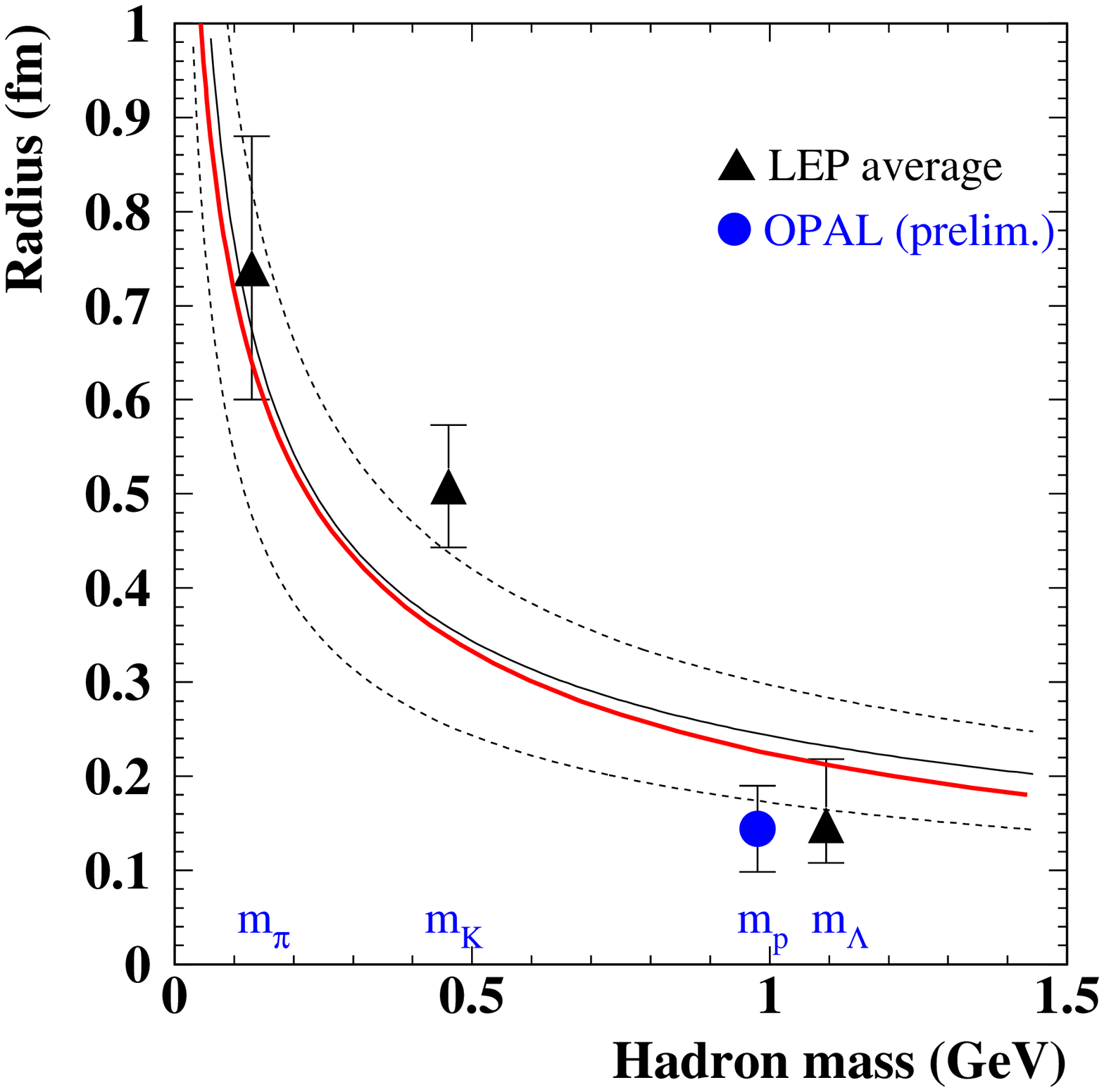}}
\caption{\label{fig:AlexanderMt}
One-dimensional Gaussian radii for source of various particles from $e^+-e^-$ collisions at LEP.
From~\cite{Alexander:2003ug}
}
\end{minipage}
\end{figure}

Thus, both the rapidity and the $m_T$ dependence in the data support the common assumption of
boost-invariance built into several parameterized models used to study transverse flow, discussed next
in Section~\ref{sec:dataTransverseFlow}.

%Finally, we note that the Gaussian radius $R_{long}$ measures $\tau$, the duration of the collision evolution until freezeout,
%according to Equation~\ref{eq:rootmT}; {\bf this is discussed} in Section~\ref{sec:Scale};
%In contrast, the non-Gaussian tail (Equation~\ref{eq:mT} is driven by thermal smearing in 
%the longitudinal direction; {\bf this is discussed} in Section~\ref{sec:halo}.

\subsection{Transverse Flow: $m_T$ Systematics and Non-identical Particle Correlations}
\label{sec:dataTransverseFlow}

Several parameterizations have been applied to HBT data which are based on thermal emission from a radially and 
longitudinally expanding source \cite{Retiere:2003kf,Csorgo:1995bi,Helgesson:1997zz,Kisiel:2006is,Kisiel:2008ws}. 
The term ``blast wave'' refers to any number of parameterizations which usually include a transverse radius, a 
maximum transverse velocity at the surface, a breakup time (which in a Bjorken geometry determines the longitudinal 
velocity gradient, $dv/dz=1/\tau_b$), and a temperature and chemical potential. Parameterizations might or might not 
include a surface-diffuseness or an emission duration.  Whereas most parameterizations assume a step-function profile, 
the Buda-Lund parameterization has a Gaussian shape and can be motivated by exact solutions to a specific solution for
non-relativistic hydrodynamics~\cite{Csorgo:1995bi,Helgesson:1997zz}.  Femtoscopic signatures of transverse flow
from these parameterizations are rather generic.  The consistency of the longitudinal systematics with an approximately
boost-invariance scenario, discussed in the previous Section, motivates the assumption of such a scenario in most models.

The $p_t$ dependence of $R_{\rm side}$ and $R_{\rm out}$ is generated by transverse flow\footnote{For sources of order 1~fm,
some of the $p_T$ dependence of $R_{\rm out}$ is also generated by the resonance halo.}, and does not necessarily 
follow any scaling formula. For instance, if one were to consider an infinite Hubble flow, all three radius 
parameters would be independent of momentum (if expressed in the pair frame) and would be equal to $\tau\sqrt{T/m}$. 
However, if one cuts off the Hubble flow at a given radial size $R_{\rm max}$ and confines emission within that 
volume, the effective source sizes are more strongly reduced for high momentum particles, for whom a larger fraction 
of the emission region would otherwise reside outside $R_{\rm max}$.  The effect is indicated by the smaller homogeneity
regions for faster (or higher $m_T$) particles in Figure~\ref{fig:blastcartoon}.

\begin{figure}[t]
\begin{minipage}[t]{0.42\textwidth}
\centerline{\includegraphics[width=0.95\textwidth]{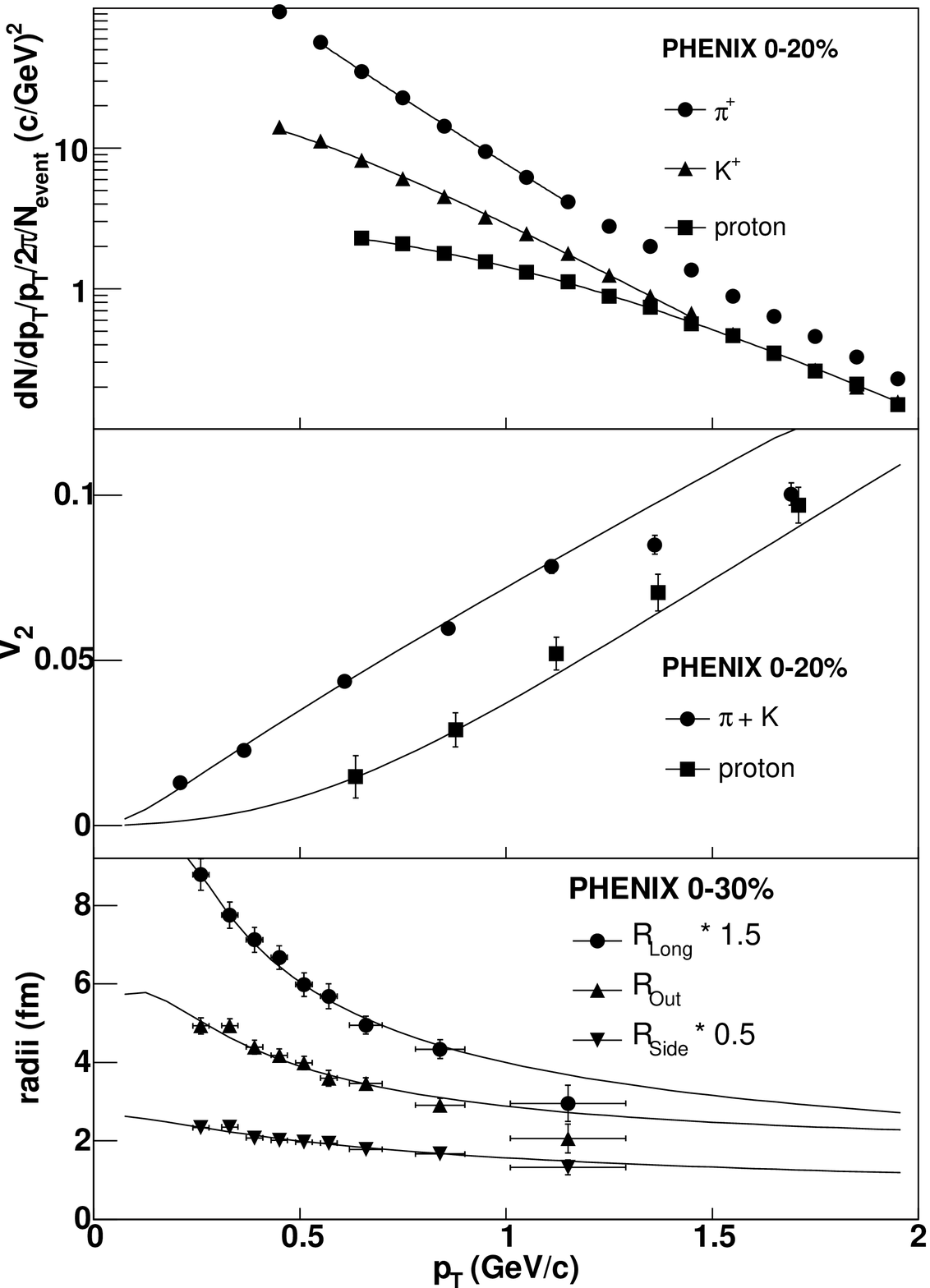}}
\caption{\label{fig:BWfit}
Blast-wave~\cite{Retiere:2003kf} calculations are compared to single-particle spectra, elliptic flow, and Gaussian ``HBT radii''
from Au+Au collisions at RHIC.
Data are plotted as a function of transverse momentum relative to the beam direction.
(Figure from \cite{Retiere:2004wa})
}
\end{minipage}
\hspace{\fill}
\begin{minipage}[t]{0.54\textwidth}
\centerline{\includegraphics[width=0.95\textwidth]{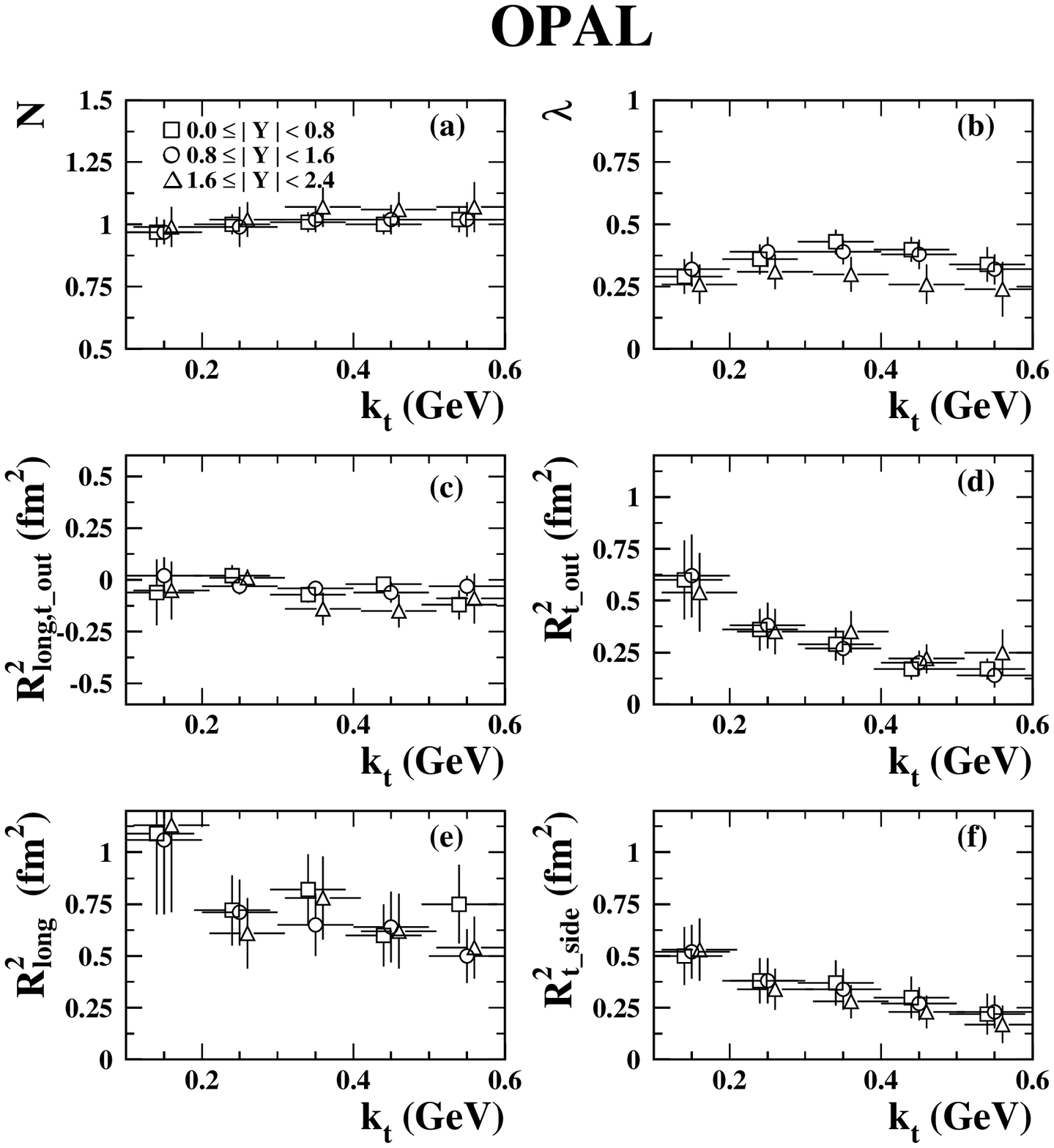}}
\caption{\label{fig:OPALkT}
Gaussian ``HBT radii'' from $e^+-e^-$ collisions at LEP, as a function of pair transverse momentum
relative to the thrust direction.
From~\cite{Abbiendi:2007he}.
}
\end{minipage}
\end{figure}

The femtoscopic flow signature is well-described by parametric models which also quantitatively reproduce momentum-only observables
such as $p_T$ spectra for identified particles and elliptic flow, as seen in Figure~\ref{fig:BWfit}.  Including realistic resonance
contributions varies the parameters (e.g. temperature) of the source, but not the requirement of a flow-dominated scenario~\cite{Kisiel:2006is}.
The femtoscopic flow signature is, in fact, observed over a huge range of conditions~\cite{Lisa:2005dd}.
System timescales are observed to grow with bombarding energy~\cite{Alt:2007uj} and collision centralities~\cite{Adams:2004yc}.
In all cases, freeze-out temperatures of order 120~MeV and flow velocities on order of 0.5c are indicated by the kinematic
dependences of the source radii.%%%%%%%%; this approximately ``universal'' behavior is discussed further in Section~\ref{sec:overall}.

\begin{table}[t]
\centering{
%%%\begin{tabular}{|p{4mm}||p{16mm}|p{16mm}|p{5mm}|p{5mm}|p{5mm}|p{11mm}|p{7mm}|p{4mm}|p{4mm}|p{4mm}|p{4mm}|}
%%\begin{tabular}{|p{4mm}||p{12mm}|p{12mm}|p{12mm}|p{12mm}|p{12mm}|p{12mm}|p{12mm}|p{12mm}|p{12mm}|p{12mm}|p{12mm}|}
\begin{tabular}{|p{4mm}||p{12mm}|p{12mm}|p{12mm}|p{12mm}|p{12mm}|p{12mm}|p{12mm}|}
%%%%%%%\begin{tabular}{|c||c|c|c|c|c|c|c|c|c|c|c|}
\hhline{--------}
~ & $\pi^+$ & $\pi^-$ & $K^+$ & $K^-$ & $K^0_s$ & $p$ & $\bar{p}$\\
\hhline{========}
$\bar{\Xi}$       & \cite{Chaloupka:2006sj}   & \cite{Chaloupka:2006sj}   &    &    &    &    &       \\ \hhline{--------}
$\Xi$             & \cite{Chaloupka:2006sj}   & \cite{Chaloupka:2006sj}   &    &    &    &    &        \\  \hhline{--------}
$\bar{\Lambda}$   &    &    &    &    &    &  \cite{Adams:2005ws}  & \cite{Adams:2005ws}       \\  \hhline{--------}
$\Lambda$         &    &    &    &    &    &  \cite{Adams:2005ws}  & \cite{Adams:2005ws}       \\  \hhline{--------}
$\bar{p}$         & \cite{Kisiel:2006si}   & \cite{Kisiel:2006si}   & \cite{Kisiel:2006si}   & \cite{Kisiel:2006si}   &    & \cite{Gos:2006ry}   & \cite{Gos:2006ry}   \\  \hhline{--------}
$p$               & \cite{Kisiel:2006si}   & \cite{Kisiel:2006si}   & \cite{Kisiel:2006si}   & \cite{Kisiel:2006si}   &    & \cite{Gos:2006ry,Heffner:2004js}   \\  \hhline{-------}
$K^0_s$           &    &    &    &    &  \cite{Bekele:2004ci}  \\  \hhline{------}
$K^-$             & \cite{Adams:2003qa}   & \cite{Adams:2003qa}   &    & \cite{Heffner:2004js}   \\  \hhline{-----}
$K^+$             & \cite{Adams:2003qa}   & \cite{Adams:2003qa}   & \cite{Heffner:2004js}   \\  \hhline{----}
$\pi^-$           & \cite{Adams:2004yc}   &   \cite{Adler:2004rq,Back:2004ug} \\  \hhline{---}
$\pi^+$           & \cite{Adler:2004rq,Back:2004ug}   \\  \hhline{--}

\end{tabular}
}        %%%%%%% this ends the ``centering'' environment
\caption{A table of published or ongoing femtoscopic studies at RHIC for various particle combinations.
         ``Traditional'' identical-particle interferometry lies along the lowest diagonal line of cells.
         The table lists representative reports only; it is not exhaustive.
\label{tab:nonid}}
\end{table}

A further test of the flow scenario is to look at other particle types, in addition to pions.
Homogeneity regions have been probed for the most common particle types.  Table~\ref{tab:nonid} indicates some of the
studies reported at RHIC.
A flowing system would produce homogeneity lengths following an approximately universal dependence on $m_T$, independent
of particle type.  As seen in Figure~\ref{fig:heavyIonMt}, a universal scaling is, in fact observed in heavy ion collisions.
A similar mass dependence is shown in Figure~\ref{fig:AlexanderMt} for lepton collisions at LEP.

Probably the most convincing test of transverse flow is its generic signature of the new information
accessible in non-identical particle correlations, discussed in Section~\ref{sec:primer}.  
In particular,
as indicated in Figure~\ref{fig:blastcartoon}, not only do higher-$m_T$ particles come from a smaller homogeneity
region than low-$m_T$ ones, but their average emission point lies further away in the ``out'' direction~\cite{Retiere:2003kf}.
Few published data on non-identical particle correlations at RHIC are available, but STAR has reported 
the offset between pions and kaons~\cite{Adams:2003qa} to be between 4 and~7.5 fm, consistent with blast 
wave expectations of 7~fm~\cite{Retiere:2003kf}.
Several measurements
of non-identical particle correlations are underway at RHIC; they appear in the non-diagonal elements of Table~\ref{tab:nonid}.

\subsection{Overall Source Size and Shape}
\label{sec:size}

Collective and random motion (parameterized by flow and temperature, respectively) determine the femtoscopic substructure of the
system created in heavy ion collisions.  However, they do not determine the {\it overall} space-time scale
of the freeze-out configuration.  Consider two ``blast-wave'' systems, with identical
temperatures and flow; the second system is identical to the first, but scaled down in space-time.  They would produce identical
momentum-space observables, spectra and elliptic flow.  Femtoscopically, the $p_T$, $y$, $\phi$ and non-identical particle systematics
would be identical, {\it except} for an overall factor of two scaling in extracted length scales (``HBT radii'' etc.).  What determines
the overall length scale?

\begin{figure}[t]
\centerline{\includegraphics[width=0.55\textwidth]{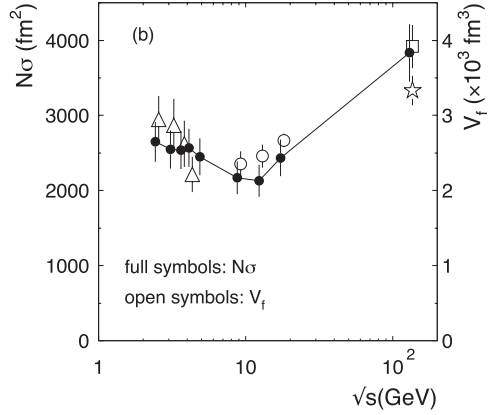}}
\caption{\label{fig:CERES}
``Effective pion cross-section'' (left axis) and pion freezeout volume (right axis) as
a function of collision energy, for central heavy ion collisions.
From~\cite{Adamova:2002ff}.
}
\end{figure}

As shown in Figure~\ref{fig:CERES}, the CERES Collaboration studied the energy 
evolution of $R_{long}R_{side}^2$ for low $p_T$ pions from central collisions between the heaviest nuclei; this quantity
corresponds roughly to the overall freezeout ``volume.''
Assuming a universal mean free path at freeze-out, they found~\cite{Adamova:2002ff} that this ``volume'' 
seemed to be driven solely by the final-state chemistry-- i.e. the yields of pions, protons, etc-- of the system.
  Above AGS energies 
($\sqrt{s_{NN}}\gtrsim 5$~GeV), the system is pion-dominated, so the determining quantity is $dN_{\pi}/d\eta\approx dN_{ch}/d\eta$.

\begin{figure}[t]
\newcommand\HR{\rule{.5em}{0.4pt}}
%%%%%%%%%\begin{figure}
\begin{minipage}[t][][t]{0.35\textwidth}
\centerline{\includegraphics[width=0.95\textwidth]{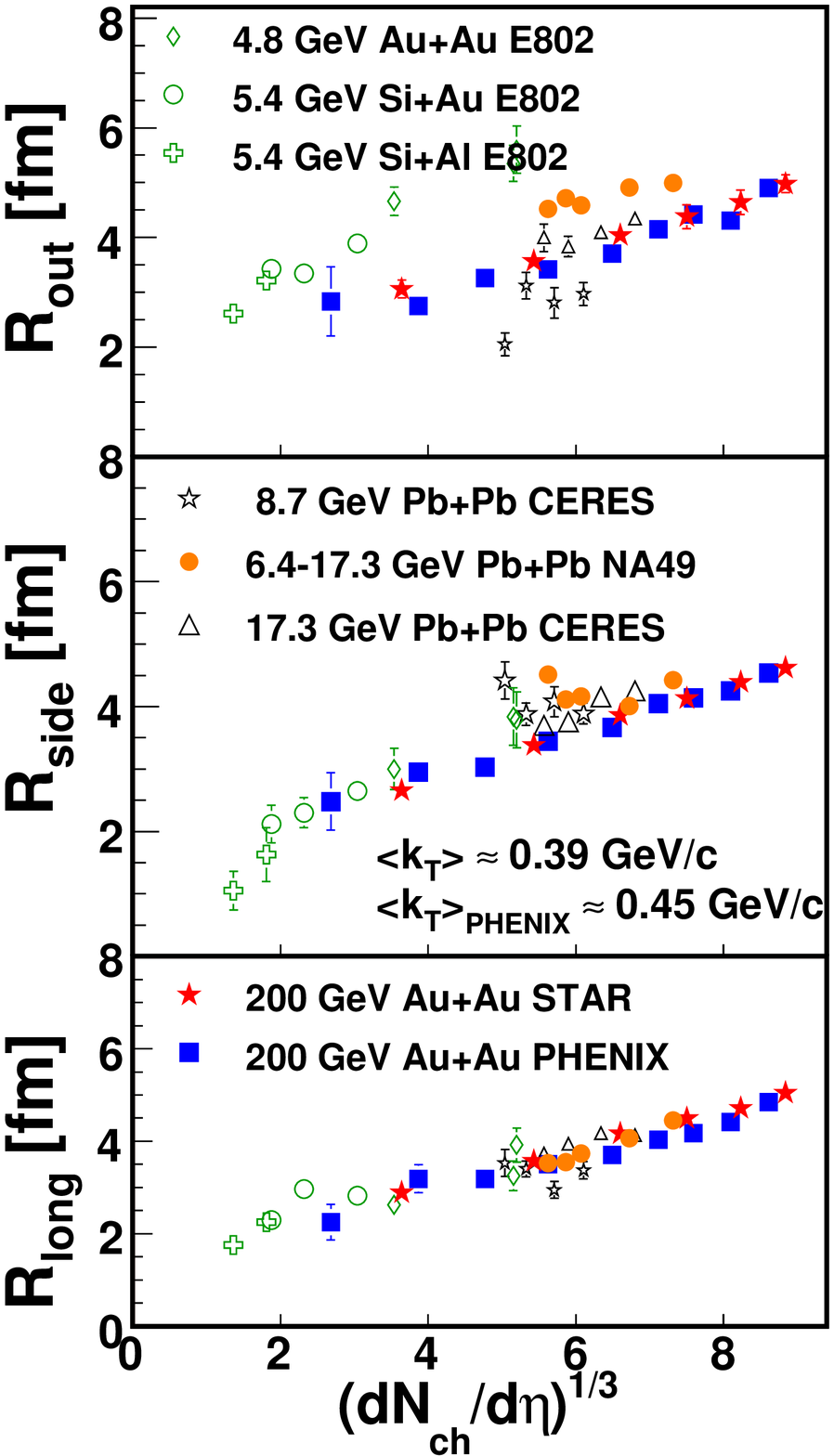}}
\caption{\label{fig:AAmultScaling}
Three-dimensional ``HBT radii'' for pions from central heavy ion collisions over a broad range of energy,
plotted as a function of charged-particle multiplicity.
}
\end{minipage}
\hspace{\fill}
%%%%%%%%\begin{minipage}[t][][t]{0.51\textwidth}
\begin{minipage}[t]{0.6\textwidth}
%%%\centerline{\includegraphics[width=0.95\textwidth]{Figures/E735multDep}}
\centerline{\includegraphics[width=0.95\textwidth]{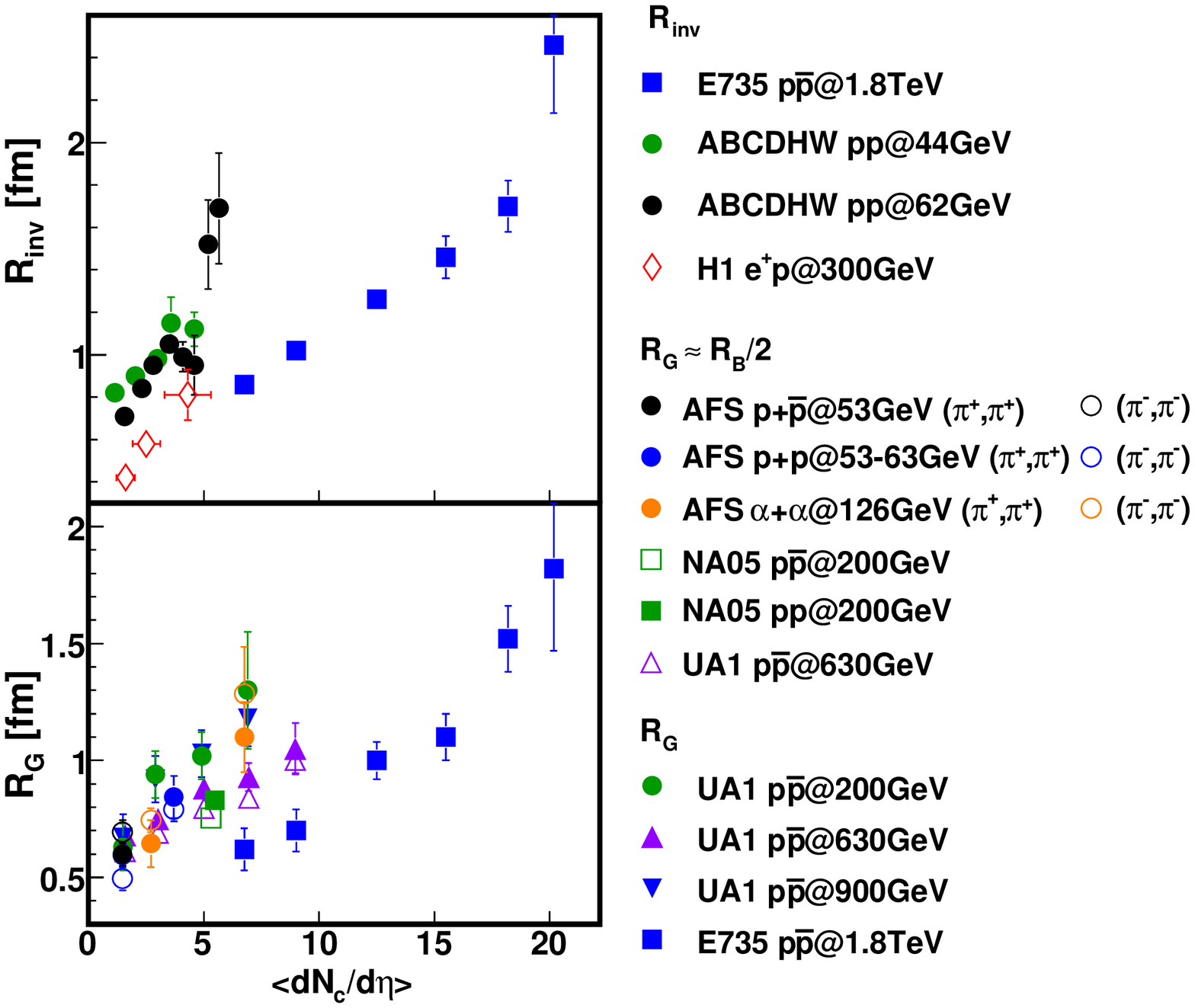}}
\caption{\label{fig:ChajeckiCompilation}
``HBT radii'' for pions from $\overline{p}-p$, $p-p$, and $\alpha-\alpha$ collisions
at the Tevatron, Sp$\overline{\rm p}$S and ISR, as a function of multiplicity.
From a compilation in~\cite{Chajecki:2009zg}.
%collisions at the Tevatron~\cite{Alexopoulos:1992iv}, 
%Sp${\overline{\rm p}}$S~\cite{Albajar:1989sj} and ISR~\cite{Akesson:1986ix}
%facilities, as a function of event multiplicity.
%From~\cite{Alexopoulos:1992iv}.
}
\end{minipage}
\end{figure}

Actually, the statement can be made much more generally.  In Figure~\ref{fig:AAmultScaling}, we see that both $R_{side}$ and $R_{long}$, individually,
scale with $dN_{ch}/d\eta$.  $R_{side}$ is related most directly to geometrical length, and $R_{long}$ to timescale~\cite{Wiedemann:1999qn,Retiere:2003kf}.
$R_{out}$ depends on both space and time in a more convoluted way~\cite{Wiedemann:1999qn,Retiere:2003kf}; 
consequently, the multiplicity scaling is broken for this HBT radius.
Since it includes results from a range of centralities in each collision system, Figure~\ref{fig:AAmultScaling} makes clear that the
chemistry-scaling is not restricted to central collisions.  Going further, the universal $p_T$-dependence mentioned
in Section~\ref{sec:dataTransverseFlow} implies that the chemistry-scaling is not restricted to any given momentum region.
Finally, inclusion of Si-induced collisions on Figure~\ref{fig:AAmultScaling} shows that
the scaling is not restricted only to collisions between the heaviest nuclei; indeed, preliminary data~\cite{Chajecki:2005zw,Das:2007ny}
from the STAR Collaboration on Cu+Cu and Au+Au collisions at various energies,
and even from d+Au and p+p collisions, further support this conclusion.

As seen in the compilation~\cite{Chajecki:2009zg} of Figure~\ref{fig:ChajeckiCompilation}, a similar multiplicity dependence
is observed in high energy hadron collisions.  Since the measurements shown were performed with somewhat different
analysis techniques and momentum acceptance, comparing the absolute magnitude of the radius experiment-to-experiment 
is nontrial.  However, within a given experiment, the multiplicity dependence is clear.

\subsection{Anisotropic freezeout configuration: systematics relative to the reaction plane}
\label{sec:asHBT}

Overall particle yields, rapidity distributions, $p_T$ spectra and elliptic flow ($v_2$) are all
measurements of the {\it number} of particles of a given momentum.  So far, we have seen a considerable
correspondence between these observables and femtoscopic ones.

Fully integrated number yields are typically used to estimate the temperature and chemical potentials
in a chemical equilibrium scenario.  Meanwhile, the overall femtoscopic scale (or volume) is driven also by the chemical
composition at freezeout.  Similarly, the shape of $dN/d\eta$ reflects the degree of boost-invariance of the
system, as do the longitudinal systematics of femtoscopic scales.  Transverse momentum spectra carry information
on explosive transverse flow, as do the $p_T$ systematics of femtoscopy scales.

By far, the most intensely studied soft-sector observable at RHIC is elliptic flow-- the number distribution
relative to the reaction plane~\cite{Voloshin:2008dg}.  Far more than the azimuthally-averaged
distribution, the anisotropic part of the momentum distribution sensitively probes important details in transport
models, such as the Equation of State~\citep[e.g.][]{Huovinen:2001cy,Kolb:2003dz}.
So far, corresponding femtoscopic studies are few,
but they may be as important as elliptic flow measurements.  Azimuthally-sensitive femtoscopy reveals the anisotropic
shape of the emitting source at freezeout.  Knowing the strength of the anisotropic flow and the initial geometric
anisotropy from a Glauber calculation, one may estimate the time required for the system to evolve from its
initial to final shape.  Such estimates~\cite{Lisa:2003ze} yield evolution timescales consistent with those
estimated by $R_{long}$ (c.f. Section~\ref{sec:longitudinalFlow}).

\begin{figure}[t!]
\centerline{\includegraphics[width=0.75\textwidth]{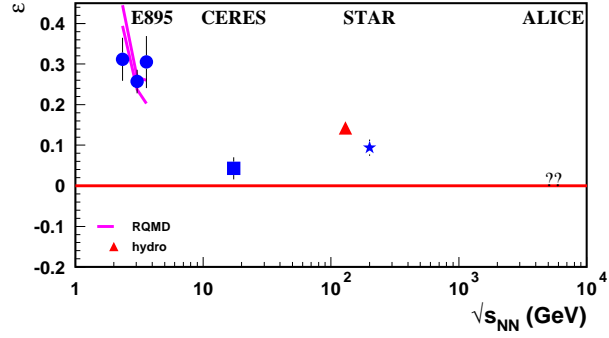}}
\caption{\label{fig:asHBTdata}
The transverse spatial freezeout anisotropy $\epsilon$ as a function of collision energy, for midcentral ($\sim 10-30\%$)
heavy ion collisions.  Round sources correspond to $\epsilon=0$; $\epsilon>0$ indicates an out-of-plane-extended source.
Measurements at the AGS~\cite{Lisa:2000xj}, SPS~\cite{Adamova:2008hs} and RHIC~\cite{Adams:2003ra} 
are compared with RQMD transport model calculations at the low energies; hydrodynamic
calculations~\cite{Heinz:2002sq} at RHIC and the LHC are indicated.
}
\end{figure}

Whereas the overall magnitude and the $p_T$- and $y$-systematics of femtoscopic scales vary little with bombarding energy~\cite{Lisa:2005dd},
the anisotropic shape varies considerably.  Figure~\ref{fig:asHBTdata} shows the world dataset of azimuthally-sensitive femtosopy measurements.
The source anisotropy $\epsilon$ is defined and estimated~\cite{Retiere:2003kf} according to
\begin{equation}
\epsilon \equiv \frac{R^2_y-R^2_x}{R^2_y+R^2_x}\approx 2\frac{R^2_{s,2}}{R^2_{s,0}} ,
\end{equation}
where $R_x$ ($R_y$) represent source dimensions in (out of) the reaction plane.  $R^2_{s,0}$ and $R^2_{s,2}$ are the
azimuthally average of the $R_{side}$ ``HBT radius'' (c.f. Section~\ref{sec:GaussianParameterizations}) and the strength
of its oscillation with azimuthal angle.

At the AGS, the emission zone is almost as anisotropic as the nuclear overlap region~\cite{Lisa:2000xj}, not surprising since elliptic flow vanishes
at these energies~\cite{Pinkenburg:1999ya}.  As both the evolution duration and strength of anisotropic flow increase with $\sqrt{s_{NN}}$, the source
becomes more round, and is even expected to become in-plane-extended at the LHC, according to hydrodynamical calculations~\cite{Heinz:2002sq,Kisiel:2008ws}.
Taken at face value, the data suggest a non-monotonic energy dependence, perhaps with a dip at SPS energies, where there are other hints
of threshold behavior~\cite{Gazdzicki:2004ef}.  However, much more systematic data is needed to confirm this intriguing energy dependence.

\subsection{Resonance decay contribution: non-Gaussian ``imaging'' fits}
\label{sec:halo}

\begin{figure}
\centerline{\includegraphics[width=0.45\textwidth]{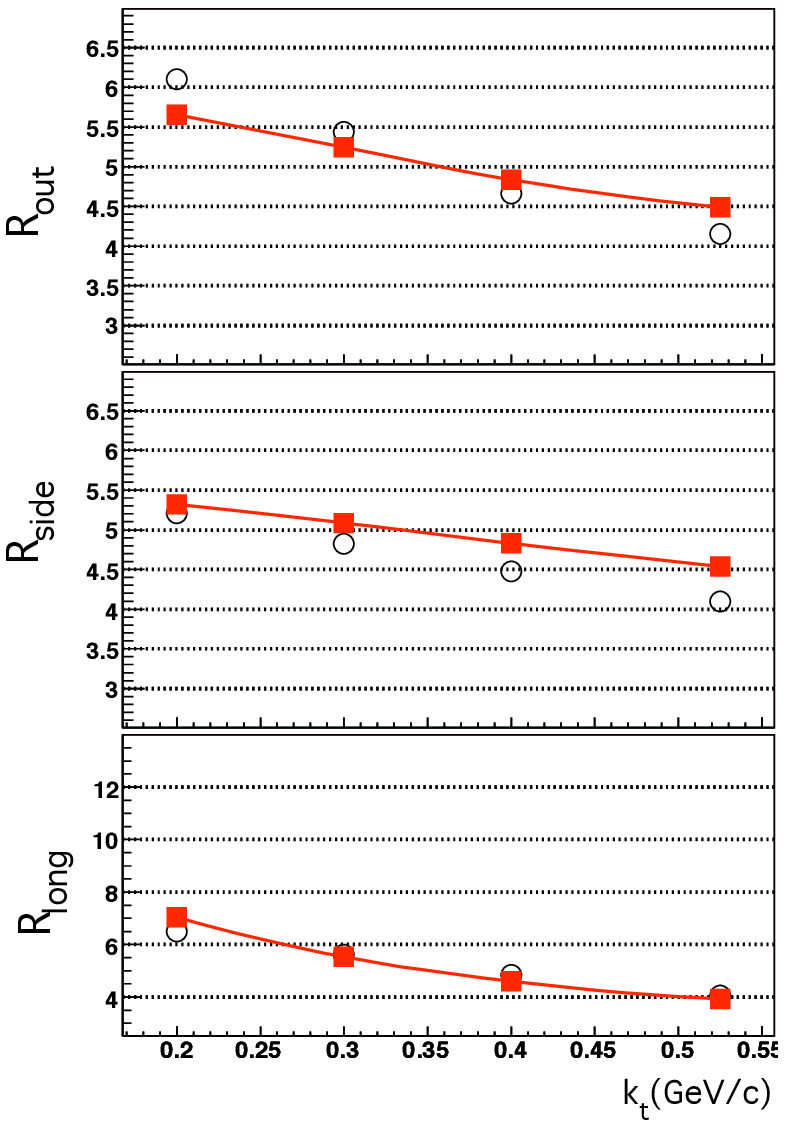}
~~\includegraphics[width=0.45\textwidth]{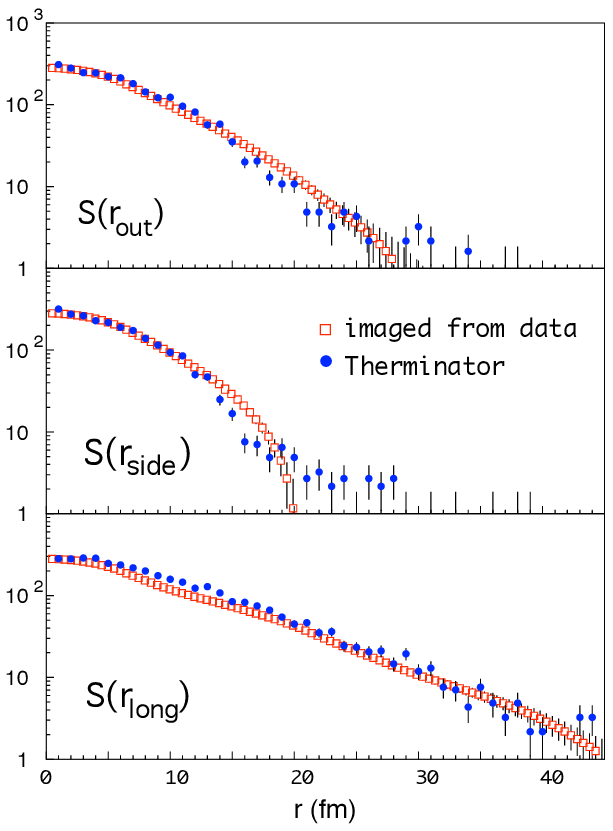}}
%%%%%%{Figures/na49_fig4_ppg}}
\caption{\label{fig:therminator}
Left panel: Gaussian source radii (in fm) as determined by fits to STAR data (circles) are fit by therminator (squares), a blast-wave model including resonant decays.\newline
Right panel: Source functions from therminator are compared to imaged source functions generated by inverting NA49 correlation functions.}
\end{figure}

While Gaussian radii from fits to correlation functions capture the bulk spatial scales, the emitting system
clearly has a richer structure.  In particular, one expects an exponential component to the source distribution 
(a ``halo''~\cite{Csorgo:1994in}) due to long- and medium-lived resonance decays.
As discussed in Section~\ref{sec:nonGauss}, in ``imaging'' fits~\cite{Brown:1997ku,Brown:2000aj}, a spline-based separation 
distribution $S(r)$ is 
varied such that the calculated moments of the correlation function (c.f. Equation~\ref{eq:imageCS})
match the measured ones.

Separation distributions so extracted are shown in Figure~\ref{fig:therminator}, from two-pion correlations from
central Pb+Pb collisions at the SPS~\cite{Alt:2008fq}.  Exponential-like tails arise in all three dimensions, and are due mainly to 
resonant decays, but along the longitudinal direction are also influenced by the exponential tail coming from boost-invariant 
dynamics as illustrated by Equation~\ref{eq:slong}.
The {\it Therminator} variant of the blast wave model also incorporates resonant decays \cite{Kisiel:2006is,Kisiel:2008ws}, 
and represents the state of the art in such models. On the left side of Figure~\ref{fig:therminator} is shown Therminator
calculations of the Gaussian ``HBT radii,'' compared to STAR measurements.

Many fewer resonances are created at the much lower energies ($\sqrt{s_{NN}}\sim 3$~GeV) used in AGS
experiments.  Thus, it is not surprising that the long-range exponential tail of $S(r)$ is not observed there~\cite{Chung:2002vk}.

\subsection{Putting it all together: factorization and the overall picture}
\label{sec:Factorization}

The decades-long femtoscopic program in heavy ion collisions has systematically turned most of the available ``knobs'' 
to map out the landscape.  These include: initial-state conditions, such as ion size and collision centrality and energy; 
kinematic knobs such as pair rapidity and azimuthal angle; and femtoscopy between different particle types.  Overall, 
the data show a remarkable ``factorization.''  Denoting a generic femtoscopic observable (e.g. $R_{long}$) by $\gimel$,
all data indicate, approximately
\begin{align}
\label{eq:Factorization}
& \gimel\left(\sqrt{s_{NN}},A,B,|\vec{b}|,\phi,y,m_T,m_1,m_2\right) \approx \\
         & \qquad\qquad R_{g}\left(dN_{\pi}/d\eta,dN_{p}/d\eta,\dots\right)\cdot F_{a}\left(\phi,\dots\right)\cdot F_{k}\left(y,m_T,m_1,m_2\right) , \nonumber
\end{align}
where $A$ and $B$ represent the size of the colliding nuclei, $m_1$ and $m_2$ are the masses of the two particles being correlated,
and $dN_i/d\eta$ is the rapidity density of particle $i$.

According to ``Equation''~\ref{eq:Factorization}, the overall, dimensionful scale of the system is determined by
the collision conditions $\sqrt{s_{NN}}$, $A$, $B$ and $|\vec{b}|$.  Going further, there is a ``universality'' in
this dependence: it is not these initial-state conditions per se, but
the final chemistry of the collision ($dN_i/d\eta$) which determines the overall factor $R_g$.

The transverse and longitudinal collective dynamics of the system is characterized by the dimensionless function $F_k$,
which essentially does {\it not} depend on collision conditions or chemistry.  Blast-wave type collectivity leaves unmistakable
signatures on this function, which is supported by all observations.  Indeed, as suggested by Figure~\ref{fig:heavyIonMt},
the dependence of $F_k$ on the masses of the correlated particles only appears when $m_1\neq m_2$, for which a shift in the
average emission point may occur; c.f. end of Section~\ref{sec:dataTransverseFlow}.

The azimuthal dependence of the source geometry is singled out, since its dependencies have not been fully mapped.  Clearly,
there is at least some dependence on both initial-state and kinematic variables: $\sqrt{s_{NN}}$ (e.g. Figure~\ref{fig:asHBTdata}), 
$p_T$~\cite{Adams:2003ra}, and centrality~\cite{Adams:2003ra,Adamova:2008hs}.
Given its clear connection to elliptic flow, evolution time~\cite{Lisa:2003ze}, significant energy evolution, and predicted sign
change at the LHC, further femtoscopic analyses relative to the reaction plane are of high priority.

When $\gimel=R_{out}$, the factorization of Equation~\ref{eq:Factorization} is somewhat violated, as shown in Figure~\ref{fig:AAmultScaling}.
As with the azimuthal dependence of femtoscopy, $R_{out}$ clearly mixes both space and time~\cite{Wiedemann:1999qn}, which scale somewhat differently
with initial conditions.

All available femtoscopic data point to a system dominated by collective motion:
approximate longitudinal boost-invariance, strong transverse flow of order $0.5c$,
a random kinetic component characterized by a ``temperature'' $\sim 0.1-0.15$~GeV.
The dimensional scale of the system at freezeout is  determined by a universal hadronic mean free path~$\sim 1$~fm~\cite{Adamova:2002ff};
its transverse size is typically $2\times$~the initial overlap zone~\cite{Adams:2004yc}, though feed-down from resonance decay add a long-range
exponential tail.

The timescales of the collision are short.  The ``HBT radius'' $R_{long}$ and the difference between the initial and final shape of
the source are both consistent~\cite{Lisa:2003ze} with an evolution time$\sim 10$~fm/c for central heavy ion collisions.  The duration
of particle emission is short- on order 3-4~fm/c for central collisions.  Again, resonance decays add a tail to the time distributions.

These conclusions are based largely on blast-wave fits to available data.
Although such models have numerous parameters, three things should be kept in mind.
Firstly, the parameters do represent a minimal set~\citep[c.f. e.g. ][]{Retiere:2003kf}.  Spectra as a function of mass and $p_T$ are
determined by one parameter each for flow and temperature.  Once these parameters are set, they leave unavoidable fingerprints
on the function $R_k$ from ``Equation''~\ref{eq:Factorization}; they completely determine the momentum and mass dependence
of identical and non-identical correlations alike.  These fingerprints cannot be changed by the remaining parameters, which
specify the size and timescales of the system and are reflected in the data by the function $R_g$.

Despite the remarkable success of blast-wave models, one wonders whether other factors might imitate flow. For heavy ion collisions, 
an obvious candidate is cooling. If emission occurs over a more extended time, but with a falling temperature, one would also expect 
radius parameters to fall with $p_t$. However, cooling can only be important if the duration of the emission is significant, and 
extended durations are incompatible with the experimental observation that $R_{\rm out}/R_{\rm side}\sim 1$. Thus, it becomes difficult 
to offer any plausible counter-scenario, to compete with the picture of a collision dominated by a rapidly developed collective radial 
flow, followed by a sudden dissolution.

\subsection{What about small systems?}
\label{sec:TooUniversal}

The study of ultrarelativistic heavy ion- as opposed to hadron or lepton-- collisions is driven by the
desire to create a bulk, self-interacting {\it system} whose properties and response may eventually
be linked to an equation of state (EoS).  The EoS is not characteristic of collisions, per se, but is
a more fundamental object, directly reflecting the symmetries and properties of non-perturbative QCD.
The use of heavy ions is based on their size-- a ``large'' system (whose scale is much more than
the mean free paths of particulate degrees of freedom) is more likely to equilibrate.

The generation of a deconfined state of matter-- in which colored degrees of freedom are relevant over
length scales greater than a hadron radius-- would also seem to call for ``large'' colliding systems.
Apparently, particle collisions need not apply.

As we have discussed, femtoscopic measurements in heavy ion collisions all point to a flow-dominated
scenario consistent with all other soft-sector observables.  This flow has become both the greatest evidence
of the creation of a bulk system at RHIC and a tool to probe its nature.  Its signature (inscribed in
the function $R_k$) is seen for a huge range of initial conditions.

Indeed, as we have seen in Figures~\ref{fig:OPALYK}, \ref{fig:AlexanderMt}, \ref{fig:OPALkT} and~\ref{fig:ChajeckiCompilation},
femtoscopic systematics-- which seem to make much sense in heavy ion collisions-- are found also in hadron-hadron
and $e^+-e^-$ collisions~\citep[c.f.][ for a compilation]{Chajecki:2009zg}.
Due to technical differences in femtoscopic studies in particle and heavy ion measurements,
direct comparisons between these measurements is difficult.  However, the first apples-to-apples comparison of
pion femtoscopy from $p+p$ and $Au+Au$ collisions at RHIC~\cite{Chajecki:2005zw}, using identical methods and detection conditions, reports
an essentially {\it identical} $p_T$-dependence of $R_k$, and clear multiplicity dependence of $R_g$, consistent with
previous measurements~\cite{Chajecki:2009zg}.  

Femtoscopic measurements are not the only soft-sector observables displaying very similar systematics in heavy
ion and particle collisions.  In hadron collisions at the highest energies, the $m_T$ distributions for heavier particles are
less steep that those for light particles~\cite{Alner:1987wb,Alexopoulos:1990hn,Levai:1991be}, 
just as they are in heavy ion collisions (c.f. Figure~\ref{fig:BWfit}).
Recent studies~\cite{Chajecki:2008yi}
suggest that
the observed breakdown of this similarity at low energies~\cite{Xu:1996xd} may be due to simple energy and momentum conservation effects.
Similarly, conservation of discrete quantum numbers (strangeness, baryon number) leads to differences in particle yield ratios measured
in heavy ion and $e^+-e^-$ collisions; but when these are accounted for, the trends are very similar~\cite{Becattini:1997rv}.
%% the following sentence was added in version 2 (10feb2009) sent to Springer
In fact, thermochemical fits to yields from $p+p$ collisions at $\sqrt{s}=200$~GeV~\cite{Abelev:2006cs} are essentially
identical to similar fits to Au+Au collisions at the same energy~\citep[e.g. Fig 12 of][]{Adams:2005dq}.

%Indeed, it may be that the only soft-sector observable which is clearly different in particle and heavy ion collisions is
%the elliptic flow signal.  Even this might be due only to the difficulty in disentangling jet contributions from azimuthal
%anisotropies of the ``underlying event'' in high-energy particle collisions.  

%Here, we do not discuss these strong similarities between particle and heavy ion collisions futher.  However,
The data obviously beg the question: if the soft sector in one collision system appears identical to that in
another, is there any firm justification to assume that the physics driving these observables is different?

In considering this question, some important points may be kept in mind.
One may imagine that string-breaking or mini-jet dynamics dominate the correlations in hadronic collisions, so
that perhaps the same mechanisms drive $R_k$ in heavy ion collisions.  However, the combined observations of Sections~\ref{sec:dataTransverseFlow}
and ~\ref{sec:size}
directly contradict this.  A minijet or string has a characteristic scale, presumably $\sim 1$~fm.  The growth of
the homogeneity region with multiplicity (e.g. Figure~\ref{fig:AAmultScaling}) might be understood in terms of 
correlations between particles from different minijets.  However in this case, there would be very little $p_T$-dependence
of $R_k$; particles with high momentum would arise from all regions of the ``whole source,'' as would slower particles~\cite{Chajecki:2005qm}.
Similarly, the suggestion~\cite{Alexander:2003ug} that the Heisenberg uncertainty principle explains the space-momentum correlation
in small systems cannot apply equally to heavy ion collisions; there is no good reason to imagine a ``scaled uncertainty principle''
in which $p_T\cdot R\approx\chi \hbar$, where $\chi$ grows to $\sim 6$ for the most central heavy ion collisions.

The striking resemblance of hadronic and nuclear collision systems
deserves much further theoretical focus and experimental exploration at RHIC and the LHC, including measurement
of non-identical particle correlations in small systems.

\section{Using Transport Models and Femtoscopy to Determine Bulk Properties}
\label{sec:bulk}

Any reasonable inference of bulk properties of matter such as the equation of state or viscosity from data requires detailed analysis with sophisticated models of the entire evolution. Thus one can not understand static properties of bulk matter without simultaneously understanding the dynamics. The closest thing to an exception to this rule concerns the entropy. Using a combination of femtoscopy and spectra, one can measure the net entropy per unit rapidity of the final state \cite{Pal:2003rz}. This is accomplished by first extracting the phase space density by dividing the spectra by the femtoscopic volume. If one assumes that entropy is conserved, one could infer a point on the equation of state if one knew the energy density and volume at some early time where the matter has thermalized but before transverse flow developed. At a time of $\tau=1$ fm/$c$, one can state the volume of the source with some confidence, $V=\tau\pi R^2$, where $R$ is the initial radius of the fireball. However, the net energy is less certain. The transverse energy per rapidity can be measured for the final state, but neglects the energy associated with longitudinal thermal motion, and also neglects the longtudinal work done by the expansion between $\tau=1$ fm/$c$ and breakup. Nonetheless, the final state entropy, $\sim 4450\pm 10\%$ units for central $130A$ GeV Au+Au collisions, was in the range expected from lattice gauge theory. Of course, this inference of the initial entropy is modified if there is large entropy generation or if the matter has not thermalized at $\tau=1$ fm/$c$.

When the first pion correlation measurements at the SPS were performed by NA35 in 1988, it was hoped that femtoscopy would reveal long lived sources, proving the existence of a strong first-order phase transition. The reasoning behind the expectations was based on model calculations assuming matter had completely stopped, and thus would not reaccelerate except for the pressure. For matter at the border of the mixed and QGP phases, the ratio of pressure to energy density, $P/\epsilon$, would be a minimum, and one would observe a maximum of the lifetime. For simple bag model equations of state, this ratio was expected to fall to approximately 3/40 at energy densities near 2 GeV/fm$^3$. If the matter had indeed stopped, and if the bag model equation of state were true, the system would require more than 50 fm/$c$ to dissolve \cite{Pratt:1986cc,Rischke:1996em}. 

Such dramatic lifetimes were not observed at the AGS and SPS, even though the expected energy density was attained somewhere in the lower SPS region. Two main reasons contributed to the failure. First, as expected, the stopping stage was transparent which mean the system would dissolve due to longitudinal expansion which did not rely on pressure or transverse flow. Nonetheless, one would have expected lifetimes near 20 fm/$c$ even with the transparency. The fact that it appears that dissolution times were closer to 10 fm/$c$ is mostly due to the fact that the true equation of state significantly varies from that of the simple bag model equation of state. Current lattice calculations show no first-order phase transition and show $P/\epsilon$ staying more than a factor of two above the bag-model minimum~\cite{Cheng:2007jq}. Additionally, the critical energy density is reached for the lower SPS range, where the non-negligible net baryon density which might provide extra repulsion.

\begin{figure}[h]
\centerline{\includegraphics[width=0.6\textwidth]{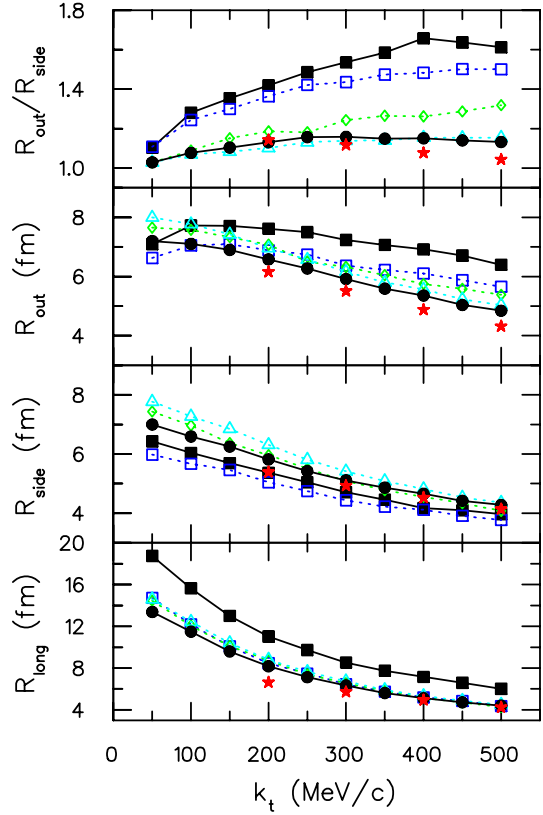}}
\caption{\label{fig:kitchensink}
Gaussian radii reflecting spatial sizes of outgoing phase space distributions in three directions: $R_{\rm out}$, $R_{\rm side}$ and $R_{\rm long}$. Data from the STAR collaboration (red stars) are poorly fit by a model with a first-order phase transition, no pre-thermal flow, and no viscosity (solid black squares). Correcting for all those deficiencies, and using a more appropriate treatment of the relative wave function in the Koonin equation \cite{Lisa:2005dd,Koonin:1977fh} brings calculations close to the data (filled black circles). The sequential effects of including pre-thermal acceleration (open blue squares), using a more realistic equation of state (open green diamonds), and adding viscosity (open cyan triangles) all make substantial improvements to fitting the data. An improved relative wave function yielded modest improvements (compare open cyan triangles to filled black circles).  See~\cite{Pratt:2008bc}
}
\end{figure}

Energy densities at RHIC surpass those of the SPS, and at 1 fm/$c$ are in the neighborhood of 10 GeV/fm$^3$, roughly a factor of 5-10 higher than those where one might expect a minimum in $P/\epsilon$. As one would expect, femtoscopic inferences of the dynamics at RHIC show somewhat more explosive collisions than at the SPS or AGS. Although the trends with beam energy were qualitatively in line with expectations, quantitative predictions with the hydrodynamic models of the early RHIC era \cite{Soff:2000eh,Kolb:2003dz}, which had been so successful in describing elliptic flow observables, provide poor reproductions of the data. The failure of the model in \cite{Soff:2000eh} was particularly disquieting given that it incorporated a cascade afterburner which should provide an accurate description of the breakup stage. In these models, the characteristic time for emission (related to $R_{\rm long}$) of particles seemed to be $\approx 15-20$ fm/$c$, rather than the $\lesssim 10$ fm/$c$ one would infer from blast wave analyses. Furthermore, the duration of emission was significantly longer in the models than in the blast-wave analyses of the data, as the models predicted $R_{\rm out}/R_{\rm side}\gtrsim 1.5$, as opposed to the experimental value of $\approx 1$. In contrast, cascade models, which many would have expected to fail given the lack of important features in the equation of state, provided a significantly better description of the data \cite{Lin:2002gc,Humanic:2002iw,Humanic:2003gs}. The effective equations of state for the cascade models was stiffer than those used in the hydrodynamic models, with some softness provided by resonances. A cascade model written without the inclusion of resonances \cite{Molnar:2002bz} significantly under-predicted experimental source sizes. 

The disparate array of models mentioned above appeared to provide a consistent trend, in that models with stiffer equations of state provided more explosive dynamics, which lead to smaller Gaussian source parameters and smaller $R_{\rm out}/R_{\rm side}$ ratios. However, one must remain wary of several significant shortcomings of the models, which might significantly affect any conclusions generated by comparison to femtoscopic data. The following list of model features and parameters have all been shown to significantly influence femtoscopic results.
\begin{itemize}
\item Stiffer equation of state: A stiffer equation of state signifies a reduced entropy, which for a fixed spectra gives smaller femtoscopic volumes. Additionally, a higher pressure leads to more rapid expansions which gives smaller $R_{\rm out}/R_{\rm side}$ ratios. Whereas calculations from earlier in the RHIC era used first-order equations of state \cite{Soff:2000eh}, incorporating a finite speed of sound, as seen in lattice calculations \cite{Cheng:2007jq}, in the region near $T_c$ can lower the $R_{\rm out}/R_{\rm side}$ ratio at the 10\% level \cite{Pratt:2008sz}.
\item Improved modeling of non-equilibrium hadronic chemistry: Significant acceleration occurs during the hadronic phase, during which chemical abundances tend to be over-populated. This lowers the pressure at the periphery which helps a pulse form which leads to a more simultaneous disassociation and an $R_{\rm out}/R_{\rm side}$ ratio which is lower by $\sim 10$\% \cite{Hirano:2002ds}.
\item Longitudinal Acceleration: Nearly all hydrodynamic codes used for femtoscopic comparison have used boost invariance, where all fluid elements have zero acceleration in the longitudinal direction, to simplify calculations. Including longitudinal acceleration breaks the simple relation between the longitudinal velocity gradient and the lifetime. This shortcoming might lead to models overstimating $R_{\rm long}$, and for parametric blast-wave analyses could lead to an underestimate of true collision lifetimes by 5-10\%\cite{Pratt:2008jj,Renk:2004kk}.
\item Shear and bulk viscosity: At early times, where the velocity gradient is large and longitudinal, shear alters the isotropy of the stress-energy tensor by lowering the effective longitudinal and increasing the transverse pressure \cite{Pratt:2006ss}. This accelerates 
the transverse acceleration and can lead to a lowering of the $R_{\rm out}/R_{\rm side}$ ratios by 10-15\% \cite{Romatschke:2007mq,Romatschke:2007jx,Pratt:2008sz}. Bulk viscosity leads to a lowering of the effective pressure for energy densities near $T_c$ \cite{Karsch:2007jc,Paech:2006st} which strengthens the pulse. However, for the periphery of the collision, the pressure is unchanged and the pulse tends to largely dissipate by breakup. The effect on source radii is then rather small \cite{Pratt:2008sz}.
\item Initial energy density profile: The profile is affected by the scheme for screening collisions. The three most often-used schemes are the wounded-nucleon model \cite{Kolb:1999it,Kolb:2000sd}, color-glass motivated prescriptions \cite{Hirano:2004en,Drescher:2007cd} and collisional scaling. Of these three prescriptions, collisional scaling  gives the most compact fireball and the wounded-nucleon model leads to the least compact. More compact fireballs are more explosive and result in a lowering of $R_{\rm out}/R_{\rm side}$ ratios by 5-10\% \cite{Pratt:2008sz,Broniowski:2008vp}.
\item Pre-equilibrium transverse acceleration: At early times, $\tau<1$ fm/$c$, the departure from equilibrium are likely so large that modifications to the stress-energy tensor can not be described by Navier-Stokes concepts. If the matter is dominated by weakly longitudinal color fields the transverse pressure is significantly higher than that of an equilibrated relativistic gas \cite{Krasnitz:2002mn,Pratt:2006ss}. This can again lower the $R_{\rm out}/R_{\rm side}$ ratio by $\sim 10$\% \cite{Pratt:2008sz}. Similar conclusions have been reached by inserting an initial velocity profile into ideal hydrodynamics \cite{Gyulassy:2007zz} and adding a strongly repulsive potential into a hadronic cascade model \cite{Li:2007yd}.
\end{itemize}

% in version 2 (10feb2009), mike has used Scott's paper ``Long slow death of HBT puzzle'' figure, so reworded this paragraph
%
%None of the items in the above list was able to account for the initial $\sim 50\%$ discrepancy in the $R_{\rm out}/R_{\rm side}$ 
%ratio, by themselves. But, by incorporating all the above features into a model, the source radii were reasonably well 
%reproduced~\cite{Pratt:2008sz}. Figure~\ref{fig:kitchensink} compares experimental source radii to an ideal hydrodynamic 
%model with a first-order phase transition and no pre-equilibrium acceleration to one that includes pre-equilibrium acceleration, 
%viscosity and a more reasonable equation of state. Unfortunately, the latter calculation ignores longitudinal acceleration, 
%whereas the ideal hydrodynamic calculation is fully three-dimensional. This might explain why the viscous calculation somewhat 
%over-predicts $R_{\rm long}$. 

None of the items in the above list was able to account for the initial $\sim 50\%$ discrepancy in the $R_{\rm out}/R_{\rm side}$ 
ratio, by themselves. But, by incorporating all the above features into a model, the source radii were reasonably well 
reproduced~\cite{Pratt:2008sz}. Figure~\ref{fig:kitchensink} compares experimental source radii to those extracted from an ideal hydrodynamic 
model with a first-order phase transition and no pre-equilibrium acceleration.  Successively correcting each of these deficiencies,
one finds increasing agreement with the data.  For more details, see~\cite{Pratt:2008bc}.

Within the last year a consensus appears to be developing~\cite[e.g.][]{Broniowski:2008vp} that a sudden impulse to the flow at early time is required to explain Gaussian source-size parameters. However, the relative contribution from shear and the non-equilibrium phase is not well known. Additionally, it is difficult to disentangle the contribution from adjusting the equation of state. It is hoped that a more global analysis, including spectra, correlations and flow observables, will be able to better distinguish and determine individual aspects of the physics.

\section{Femtoscopic Expectations at the LHC}
\label{sec:Expectations}

For soft sector observables in heavy ion physics, long-term baselines have been established over a large energy range.  Prior to first data at RHIC,
it was commonly speculated (and hoped) that large deviations from these systematics (e.g. $\pi^0/\pi^\pm$ ratios, sidewards
flow, strangeness enhancement, total multiplicity) would signal clearly the qualitatively different nature of the system created
there~\cite{Harris:1996zx}.  In femtoscopic systems, rather generic arguments led
to expectations~\cite{Rischke:1996em,Bass:1999zq} of a rapid increase, with $\sqrt{s_{NN}}$,
in the pion ``HBT radii'' $R_{out}$ and $R_{long}$, reflecting relatively long timescales of the transition
from deconfined QGP to confined hadronic matter.

Such dramatic speculations are largely absent today in anticipation of LHC collisions.
Soft-sector, global observables at RHIC are only quantitatively different than they are at lower energies.
Even in the high-$p_T$ sector, where
jet suppression and partonic energy loss measurements have generated huge excitement, energy scans
at RHIC reveal that the data indicate more of an evolution than a revolution.

The need for such systematic comparisons is not unique to RHIC, but has been a generic feature of all
heavy ion programs~\cite{Nagamiya:1988ge,Tannenbaum:2006ch}, from low-energy facilities like the NSCL (Michigan State),
to progressively higher-energy facilities at
SIS (GSI), the Bevatron/Bevalac (Berkeley Lab), AGS (Brookhaven), and SPS (CERN).  The nature of heavy ion physics
is such that little is learned through study of a single system.
Discoveries via sharp jumps {\it \`{a} la} superconductivity are not our lot.
The real science behind heavy ion measurements (at very high energies as well as at much lower ones)
lies in understanding the {\it details} of the data.

With this in mind, what may await us in heavy ion collisions at the LHC?  After discussing a simple multiplicity-based
extrapolation of existing systematics, motivated by the factorization of Section~\ref{sec:Factorization}, we summarize
predictions of Boltzmann and hydrodynamical transport calculations.  We will find that, in both cases, the much greater
explosive flow predicted at the LHC may lead to new effects which break the simple extrapolation.

\subsection{Nothing New Under the Sun (NNUS) Scenario}
\label{sec:NNUS}

As discussed in Section~\ref{sec:Factorization}, most femtoscopic systematics so far measured may be summarized 
in terms of a factorization.  While the kinematic dependences may be described by a single dimensionless function $F_k$,
there is an overall scale $R_g$ which grows roughly as the cube root of the charged particle multiplicity.  

While the
factorization is somewhat broken when $\gimel=R_{out}$ and perhaps by azimuthally-sensitve femtoscopy, the overall multiplicity
is probably
our best, zero-new-physics guide to simple extrapolation of femtoscopic trends measured over two
orders of magnitude in $\sqrt{s_{NN}}$ and from from the lightest ($p+p$) to the heaviest ($Pb+Pb$)
systems.  
Figure~\ref{fig:AAmultScaling} suggests a simple form $R_g(dN/dy) \propto \left(dN/dy\right)^{1/3}$; this ignores the finite
offset $\sim 1$~fm when extrapolating $dN/dy \rightarrow 0$, but this is negligible for high multiplicity.
This relation may reflect that a constant freezeout density drives the femtoscopic scales~\cite{Adamova:2002ff},
though this neglects any dynamic effects.
Assuming that this simple proportionality continues, then,
determining femtoscopic expectations boils down to anticipating the multiplicity at the LHC.

A naive extrapolation~\cite{Lisa:2005js,Caines:2006if} of systematics suggests that
$dN/dy$ at the LHC will be 60\% larger than that observed at RHIC.  Thus, the zeroth-order
expectation is that length scales at the LHC will be 17\% ($1.6^{1/3}=1.17$) larger than
those measured at RHIC, for all kinematic selections and particle species, according
to Equation~\ref{eq:Factorization}.

Going beyond simple extrapolation to include a physical picture, saturation-based
calculations~\cite{Kharzeev:2004if} give much higher multiplicity-- roughly triple
that at RHIC.  This leads to expectations of length scales 45\% higher than those
at RHIC.  Thus, $R_{long}$ for pions at midrapidity and low $p_T$ in central collisions
would be $1.45\times 7~{\rm fm} = 10~{\rm fm}$.

Multiplicity predictions based on Boltzmann/cascade calculations can be significantly higher yet.
Selecting two for which femtoscopic predictions also exist (Section~\ref{sec:cascade}),
A Multi-Phase Transport (AMPT) calculation~\cite{Ko:2007zzc}
and the Hadronic Rescattering Model (HRM)~\cite{Humanic:2005ye} predict $5\times$ and $7\times$ RHIC
multiplicity, respectively.  Thus, femtoscopic scales at LHC may be as much as 90\% higher
than at RHIC.  Depending on the final-state interaction which produces the two-particle
correlation function, measuring length scales of $\sim 15$~fm may challenge experimental
two-track resolutions.  For two-pion correlations, such scales are within the capabilities
of the ALICE detector~\cite{Alessandro:2006yt}.

%--------------------------------------------------------------------------------------------------
\subsection{Boltzmann Transport Calculations}
\label{sec:cascade}

\vspace*{-3mm}

More interesting than simple scaling relations are models with real physics and
dynamics, such as transport calculations.
Boltzmann/cascade transport models generally reproduce ``HBT radii'' at RHIC
better than do hydrodynamic calculations~\cite{Lisa:2005dd}.  
The reasons behind this include different 
physics in the models, a more detailed description of the kinetic freezeout,
and the use of more appropriate methods of calculating the radii~\cite{Frodermann:2006sp}.
Predictions of pion HBT radii with each 
of the transport calculations discussed in Section~\ref{sec:NNUS} reveal predictions
more subtle than the simple multiplicity-scaling discussed above.

\begin{figure}[t]
\begin{minipage}[t]{0.49\textwidth}
{\centerline{\includegraphics[width=0.95\textwidth]{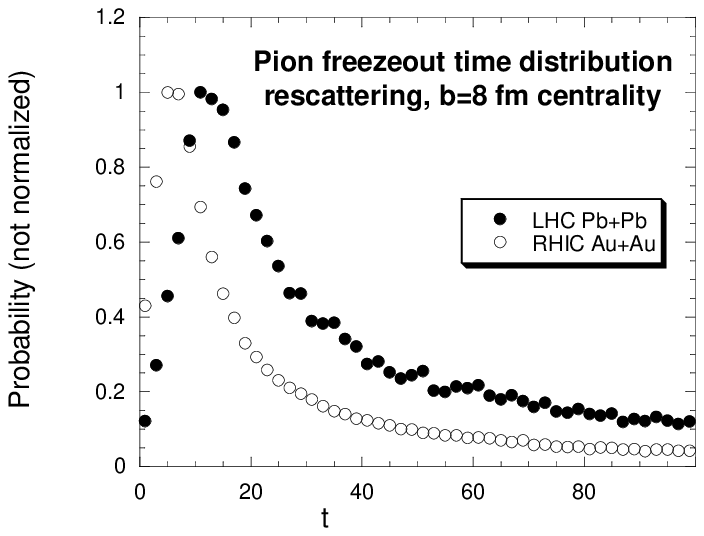}}}
\caption{
\label{fig:HRMtime}
The freezeout time distribution from the Hadronic Rescattering Model of Humanic~\cite{Humanic:2005ye} for RHIC and LHC conditions.
}
\end{minipage}
\hspace{\fill}
\begin{minipage}[t]{0.49\textwidth}
{\centerline{\includegraphics[width=0.95\textwidth]{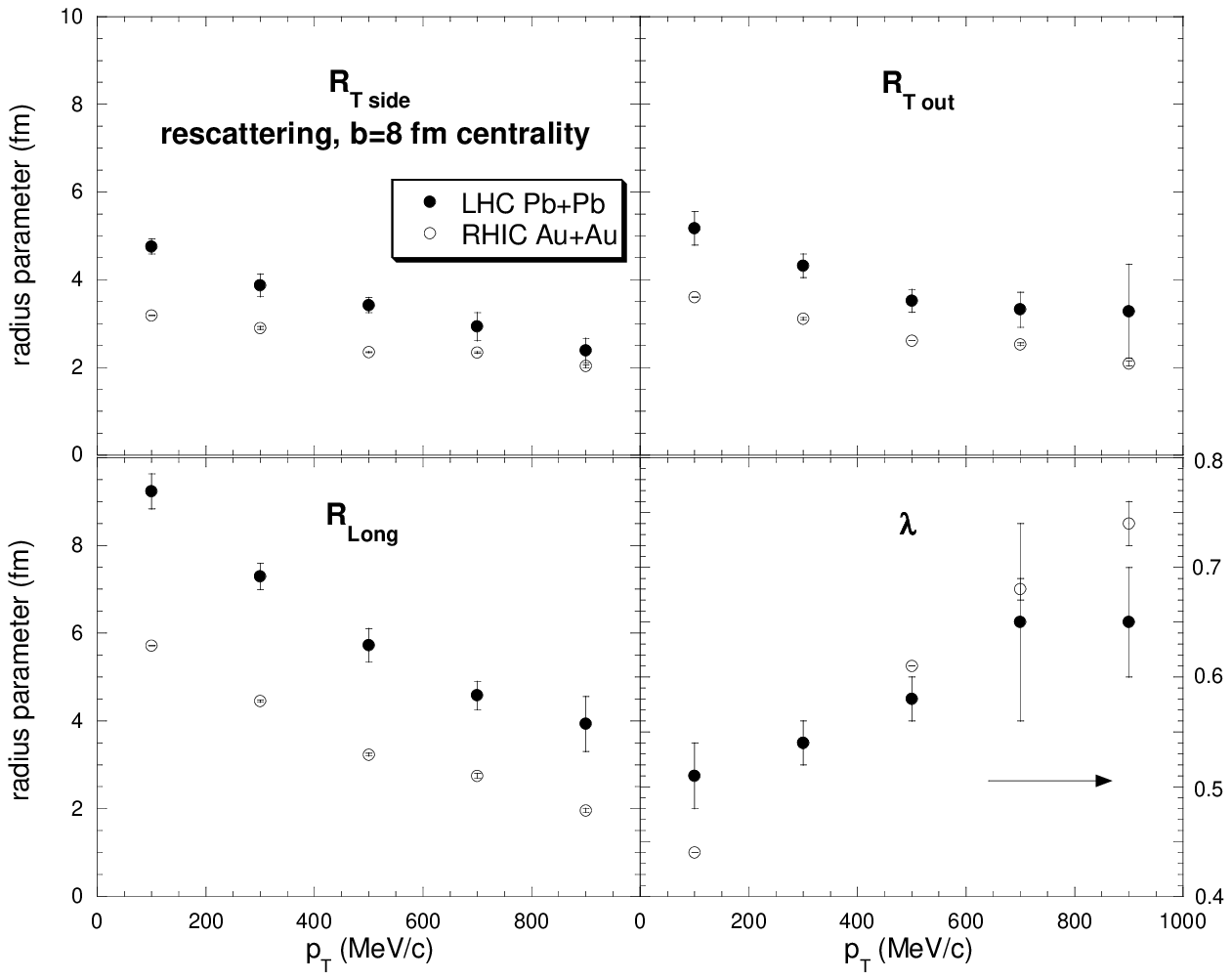}}}
\caption{
\label{fig:HRMradii}
HBT radii from fits to pion correlation functions from the
Hadronic Rescattering Model of Humanic~\cite{Humanic:2005ye} for RHIC and LHC conditions.
}
\end{minipage}
\end{figure}

For an infinite and boost-invariant system (only an approximation of reality, of course),
the longitudinal HBT radius $R_{long}$ is proportional to the system evolution
time (i.e. between interpenetration of the ions and kinematic freezeout of the
products)~\cite{Akkelin:1995gh,Lisa:2005dd}.  Naturally, this is not a unique, system-wide
time, but a distribution.  An example is seen in Figure~\ref{fig:HRMtime}, in which
the pion freeze-out time distribution for collisions at RHIC and LHC are compared in
the HRM calculation.  The LHC timescales are roughly double those at RHIC.
Although HRM is not explicitly a boost-invariant model, we see in Figure~\ref{fig:HRMradii}
that $R_{long}$ reflects this timescale increase, roughly doubling when the energy
is increased from RHIC to LHC energies.
The $\sim 70\%$ increase in $R_{long}$ is roughly consistent, then, with expectations from
both a timescale and from the multiplicity-scaling point of view.
This is certainly not a coincidence; at the LHC, the particles simply require more time to
get isolated from their more numerous neighbors.

On the other hand, there is more going on.  The predicted $p_T$-dependence of 
both $R_{long}$ and $R_{side}$ are steeper at the LHC than at RHIC.  Also, the
increase in $R_{side}$ is significantly less than 90\%.  Both of these effects
are consistent with a freezeout scenario with significantly increased transverse
flow~\cite{Retiere:2003kf}.  Indeed, transverse momentum distributions predicted by HRM
are significantly harder (less steep) at the LHC than those at RHIC.  
Since the $p_T$ dependence of HBT radii~\cite{Lisa:2005dd} and spectra  %%%%~\cite{NuXuMaybe}
in the soft sector are observed to
change very little between $\sqrt{s_{NN}} = 20 \div 200$~GeV, it will
be interesting to see whether this trend is broken at the LHC, as predicted by HRM.

The HRM model is a deliberate effort to use the simplest (often criticized as simplistic)
physics picture, free of novel phases like QGP.  It is a pure hadron-based transport calculation,
though the initial conditions may be taken from Pythia or Saturation-based scenarios~\cite{Humanic:2005ye}.
On the other side of the ``simplicity spectrum'' is AMPT, an attempt to describe the various stages of the
system's evolution in terms of the most appropriate model for that stage~\cite{Lin:2002gc}.

Similar to HRM, AMPT predicts stronger transverse flow at the LHC, as compared to RHIC, leading
to steeper $p_T$-dependence of HBT radii.  In terms of scale, the transverse (longitudinal)
radii are predicted to increase by 10\% (30\%).  This is more modest than the predictions
of HRM (30\% and 70\%, respectively), and much more modest than pure-multiplicity scalings
of Section~\ref{sec:NNUS}.

Thus the dynamical physics, in these models, lead to expected details significantly beyond
simple extrapolation of lower-energy results.

\vspace*{-3mm}

%--------------------------------------------------------------------------------------------------
\subsection{Hydrodynamical Calculations}
\label{sec:hydro}

\vspace*{-3mm}

As mentioned, hydrodynamical models tend to reproduce femtoscopic measurements
more poorly than do Boltzmann/cascade calculations.  On the other hand, they have
enjoyed huge success in reproducing momentum-space observables such as elliptic
flow.  Furthermore, the conditions at LHC are likely to provide an even better
approximation than at RHIC to the zero-mean-free-path assumptions of ideal hydrodynamics.
Finally, the direct connection between hydrodynamics and the Equation of State of
strongly-interacting matter (color-confined or not) remains a compelling reason to
explore soft-sector, bulk consequences of the model.

\vspace*{-3mm}

\subsubsection{Source Length Scales}
\label{sec:HydroScale}

\vspace*{-3mm}

Recently, Eskola and collaborators~\cite{Eskola:2005ue} coupled a pQCD+saturation-based prediction for
initial conditions at LHC to their $1+1$-dimensional hydro calculation.  The Equation-of-State
featured a first-order phase transition between an ideal QGP at high temperature and a hadron
resonance gas at low temperature.

\begin{figure}[t]
\begin{minipage}[t]{0.33\textwidth}
{\centerline{\includegraphics[width=0.95\textwidth]{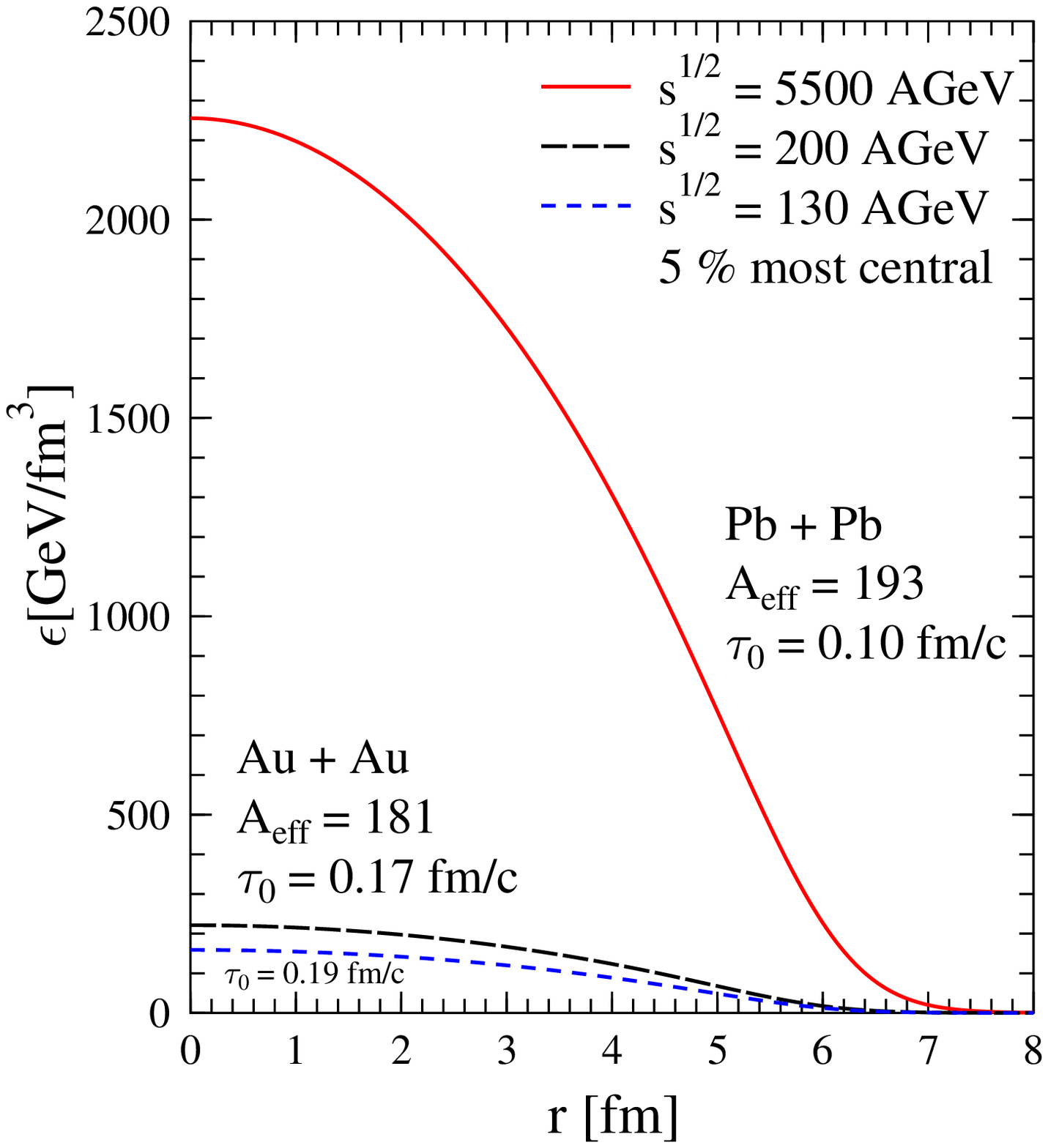}}}
\caption{
\label{fig:EskolaEnergyDensity}
Initial energy density distribution in the transverse plane, calculated by Eskola et al~\cite{Eskola:2005ue}.
}
\end{minipage}
\hspace{\fill}
\begin{minipage}[t]{0.65\textwidth}
{\centerline{\includegraphics[width=0.95\textwidth]{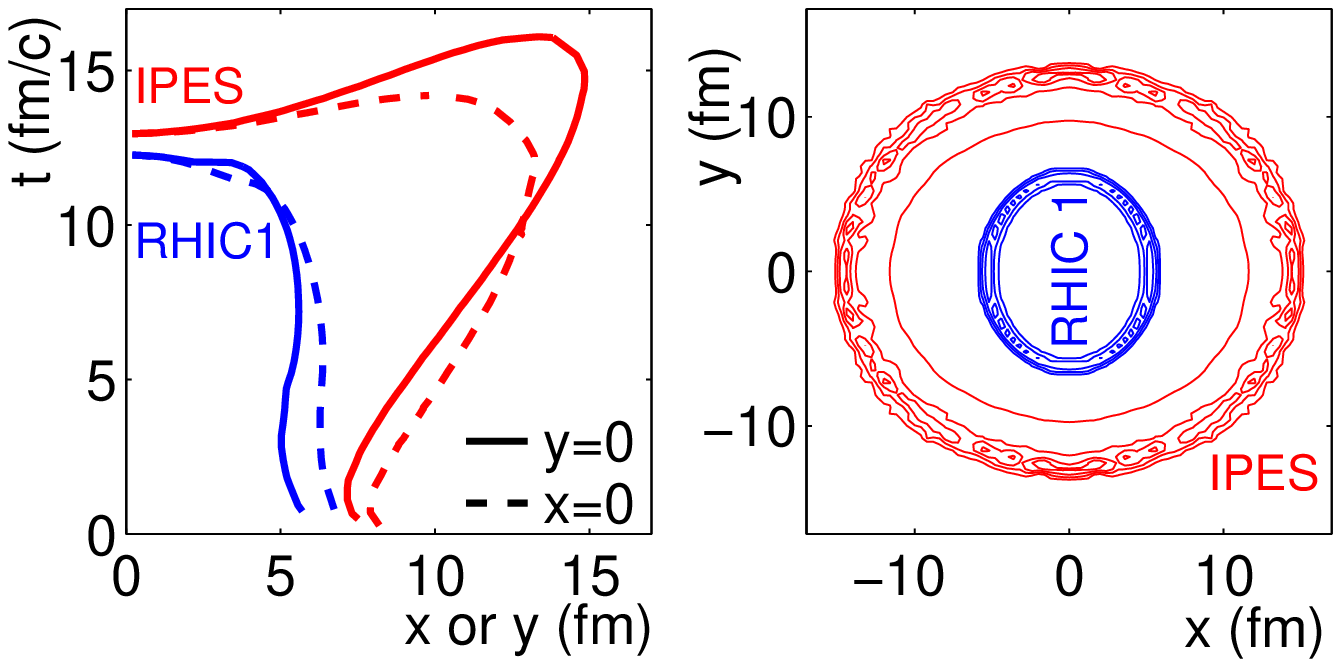}}}
\caption{
\label{fig:HeinzKolbFOhypersurface}
Left: freeze-out space-time hypersurface in collisions with finite impact
parameter calculated by Heinz and Kolb~\cite{Heinz:2002sq}.  The ``IPES'' calculation
is an estimate of the system created at the LHC.  The impact parameter is defined
to lie in the $\hat{x}$-direction.
Right: time-,z- and momentum-integrated freeze-out shapes in the transverse plane.
}
\end{minipage}
\end{figure}

As shown in Figure~\ref{fig:EskolaEnergyDensity}, the initial energy density at which hydrodynamics
is assumed to take over expected to roughly an order of magnitude larger at the LHC than at RHIC,
due both to increased gluon production and to shorter system formation (thermalization)
time $\tau_0$ at the higher energy.
Since the initial transverse scale changes only little, the pressure gradients will likewise
be higher at LHC, leading to increased transverse flow.  
These effects place competing pressures
on the space-time evolution of the system, and on the femtoscopic scales at freezeout, as discussed below.

The increased energy density (directly associated with entropy density and thus multiplicity)
tends to produce longer timescales at the LHC.  Longitudinal expansion tends to cool the system
towards freezeout conditions.  However, especially at the LHC, the large transverse flow generated
by the intense pressure gradients cannot be ignored.
Eskola and collaborators~\cite{Eskola:2005ue}, estimate that the time required to cool 
from the maximum energy density (at $r=0$ in Figure~\ref{fig:EskolaEnergyDensity} to
the critical energy density ($\epsilon_c = 1.93 GeV/fm^3$) would be 6~fm/c (20~fm/c) at RHIC
(LHC), due to longitudinal expansion alone.  However, when transverse dynamics are included,
the cooling times become 5~fm/c (7.5~fm/c) at RHIC (LHC).  The evolution time to kinematic
freezeout-- say until $T\approx 140$~MeV-- is $\tau_0\sim 12-14$~fm/c in both cases; this is the timescale most
directly probed by femtoscopy.
This is dramatic-- the effect
of transverse flow on cooling timescales can almost be neglected at RHIC, while it is dominant
at the LHC.  This is reminiscent of the cascade calculations discussed in Section~\ref{sec:cascade};
the much stronger flow may well lead to deviations from the trends (e.g. $p_T$-dependence of pion HBT
radii being independent of $\sqrt{s_{NN}}$) established so far at lower energy.  
This aspect of the NNUS scenario may finally be violated.
The qualitative difference is apparent from the freezeout hypersurfaces at RHIC and LHC, shown
in Figure~\ref{fig:HeinzKolbFOhypersurface}.  The Figure is from a
calculation by Kolb and Heinz~\cite{Heinz:2002sq}, but is similar to Eskola's.

It would be very interesting to know whether the other aspect of NNUS, namely the multiplicity
scaling shown in Figure~\ref{fig:AAmultScaling}, is satisfied by the hydro models.
Unfortunately, Eskola did not calculate pion ``HBT radii,'' and Heinz and Kolb did  not calculate
multiplicity, so a consistent estimate of the scaling cannot be checked.  However, the former
predict that the multiplicity at LHC will be approximately triple that at RHIC, corresponding
to a 40\% increase in HBT radius under NNUS scaling.
The transverse radii predicted by 
Heinz and Kolb~\cite{Heinz:2002sq} increase by a similar amount.

Recent hydro calculations by Sinyukov et al~\cite{Sinyukov:2007xa} and Chojnacki et al~\cite{Chojnacki:2007rq}
report predictions similar to those of Boltzmann calculations from the previous Section, again due to the
much larger flow.  In particular, single-particle
spectra are predicted to be much shallower at LHC than at RHIC.  
$R_{long}$ is predicted to grow by about $\sim 70\%$.  $R_{out}$ ($R_{side}$)
should increase $\sim 15\%$ ($\sim 40\%$), and fall more steeply with $p_T$ at the LHC.  These seem to be rather generic
predictions from the transport models.  They represent a deviation from the almost universal $p_T$-dependence of $R_k$ (c.f. Section~\ref{sec:Factorization})
measured thus far, and thus will be an important test of a much more explosive source predicted at the LHC.

While the hydro-based predictions Frodermann et al~\cite{Frodermann:2007ab} agree with most of those discussed above,
they are unique in that no change at all is expected for $R_{long}$.

In Table~\ref{tab:PredictedScales}, we summarize the approximate increases expected for the HBT radii at
the LHC, relative to those at RHIC.  One must keep in mind, however, that the models predict a steeper
$k_T$-dependence of the radii at the LHC, so these numbers are taken only at low $p_T$.

%%%%%%\here i am

\begin{table}[t]
\centering{
%%%\begin{tabular}{|l|c|c|c|}
\begin{tabular}{|p{35mm}|p{15mm}|p{15mm}|p{15mm}|}
\hhline{----}
~ & ~~$R_{out}$~~& ~~$R_{side}$~~ & ~~$R_{long}$~~\\
\hhline{====}
NNUS                                     & 17\%  & 17\%  & 17\%  \\ \hhline{----}
HRM~\cite{Humanic:2005ye}                & 45\%  & 45\%  & 60\%  \\ \hhline{----}
AMPT~\cite{Ko:2007zzc}                     & 10\%  & 10\%  & 30\%  \\ \hhline{----}
Eskola hydro~\cite{Eskola:2005ue}        & 40\%? & 40\%? & 40\%? \\ \hhline{----}
Heinz hydro~\cite{Frodermann:2007ab}     & 30\%  & 30\%  & ~~0\%  \\ \hhline{----}
Sinyukov hydro~\cite{Sinyukov:2007xa}    & 15\%  & 25\%  & 25\%  \\ \hhline{----}
Chojnacki hydro~\cite{Chojnacki:2007rq}  & 25\%  & 45\%  & 25\%  \\ \hhline{----}
\end{tabular}
}
\caption{Rough indications of the increase in pion HBT radii at the LHC, relative to RHIC, predicted by transport models.  Since
the models predict a stronger $k_T$ dependence at the LHC, these numbers will vary with $k_T$.  The numbers listed are taken at
low $k_T$ ($\sim 150$~MeV/c).  Our estimate for Eskola's hydro is based only on simplistic multiplicity scaling; actual HBT radii
have not been calculated in this model.
\label{tab:PredictedScales}}
\end{table}

\subsubsection{Source Shape}
\label{sec:HydroShape}

Azimuthally-sensitive pion interferometry-- the measurement of spatial scales as a function
of emission angle relative to the reaction plane-- probes the {\it shape} of the freezeout
configuration, in addition to its scale~\cite{Voloshin:1996ch,Lisa:2000ip,Retiere:2003kf}.
At finite impact parameter, both the spatial configuration of the entrance channel and the
resulting momentum distribution in the exit channel are anisotropic.  In particular, the
initial state is spatially extended out of the reaction plane, and the resulting flow is stronger
in the reaction plane (elliptic flow $v_2>0$).

Due to the preferential in-plane expansion, as the system evolves the spatial configuration 
should become increasingly in-plane extended (equivalently, decreasingly out-of-plane extended).
Thus, knowledge of the entrance-channel shape (e.g. though Glauber model calculations) and measurement
of the exit-channel shape (through femtoscopy) provide ``boundary conditions'' on the dynamical
spacetime evolution of the anisotropic system, and probe the evolution timescale.
The extracted timescale is model-dependent, requiring in principle a detailed time
evolution of the flow.  However, a simple estimate~\cite{Lisa:2003ze}
of the timescale extracted through shape measurements
and that extracted from blast-wave fits~\cite{Retiere:2003kf,Adams:2004yc} to azimuthally-integrated HBT radii
are roughly consistent.

Measurements of the freezeout shape at the AGS~\cite{Lisa:2000xj}, SPS~\cite{Adamova:2008hs} and RHIC~\cite{Wells:2002phd,Adams:2003ra}
indicate an out-of-plane-extended configuration.  Consistent with the fact that preferential in-plane
expansion (i.e. elliptic flow) is stronger at RHIC, the configuration at the higher bombarding energies
is rounder than at the AGS as shown in Figure~\ref{fig:asHBTdata}.

The anisotropic shapes and corresponding azimuthally-selected HBT radii have been calculated in Boltzmann and hydrodynamical
models.  At the AGS, the transport code RQMD~\cite{Sorge:1995dp} reproduces the overall scale~\cite{Lisa:2000no}
{\it and} the anisotropy~\cite{Lisa:2000ip,Lisa:2000xj} of the source reasonably well, as shown in Figure~\ref{fig:asHBTdata}.  At RHIC,
two groups~\cite{Heinz:2002sq,Chojnacki:2007rq} have reported that $2+1$ hydrodynamical codes
reproduce the shape quite well.  At the LHC, the hydro 
calculations predict~\cite{Heinz:2002sq,Kisiel:2008ws} a {\it qualitative} change in the freezeout distribution.  As shown in the right panel of
Figure~\ref{fig:HeinzKolbFOhypersurface} the source is expected to evolve to an {\it in-}plane configuration ($\epsilon < 0$).
This, and the increased overall size, arise due to the much larger flow at the LHC, together with the increased evolution time.  
Geometrical and dynamical effects combine non-trivially to generate the radius oscillations, but at very low $p_T$ geometry dominates, and one may
in fact measure an {\it inversion} of the sign of the oscillations~\cite{Heinz:2002sq,Frodermann:2007ab}.  To see this dramatic effect in ALICE, pion tracking with
the Inner Tracking System~\cite{Alessandro:2006yt} will be crucial.

\subsection{Proton Collisions}
\label{sec:pp}

Before heavy ions are accelerated at the LHC, proton collisions at $\sqrt{s}=1.4$~TeV will be measured.
While the thrust of the $p+p$ program is towards Higgs physics, it is well-recognized that $p+p$ collisions
serve as a valuable reference to heavy ion analyses in the ``hard'' (high-$p_T$) sector, where one looks
for the effects of the medium on particles coming from well-calibrated fundamental processes.  

Soft-sector analyses, too, should be performed for systems from the smallest to
the largest, and the results compared.  Since such analyses are assumed to measure
the {\it bulk} properties, one might well hope for qualitative differences when comparing results for
$p+p$ to $Pb+Pb$ collisions.

While pion HBT measurements have been common in both the high-energy and heavy-ion communities for
many years, a direct ``apples-to-apples'' comparison between results from $A+A$ and $p+p$ collisions
has not been possible until very recently.  The STAR Collaboration at RHIC has reported the first
direct comparison of pion HBT radii in Au+Au, Cu+Cu, d+Au, and p+p collisions, using the same detector,
same energy, identical techniques (event mixing, etc) to create the correlation function, identical
coordinate systems and identical fitting techniques~\cite{Chajecki:2005zw}.  Remarkably, Gaussian fits to the correlation
functions return ``HBT radii'' which factorize according to Equation~\ref{eq:Factorization}; i.e.
$F_k$, which quantifies the dynamically-generated substructure, is identical in $p+p$ and $A+A$ collisions.
However, the STAR data show significant non-femtoscopic structures~\cite{Chajecki:2005zw}, which must be properly
accounted for~\cite{Chajecki:2008vg} before drawing firm conclusions.
If the factorization is unchanged after a more sophisticated treatment, the physics implications might be dramatic.
We do not discuss this here, but simply observe that the NNUS scenario is a likely baseline expectation for $p+p$
at LHC.

\begin{figure}[]
{\centerline{\includegraphics[width=0.4\textwidth]{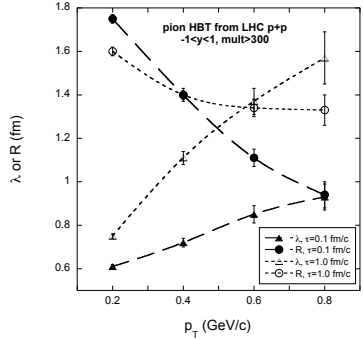}}}
\caption{
\label{fig:PythiaHRMpp}
Full markers: predicted $p_T$-dependence of $R_{inv}$ and $\lambda$ from Gaussian fits to calculated correlation
functions in a Pythia+hadronic~rescattering model~\cite{Humanic:2006ib}.  Here, the hadronization time was 
set to $\tau=0.1$~fm/c, the value required to reproduce E735~\cite{Alexopoulos:1992iv} data at the Tevatron.
Also shown in open markers are the femtoscopic parameters corresponding to a longer hadronization time.  In this
case, there is little very hadronic rescattering and thus a much weaker $p_T$-dependence of $R_{inv}$.
}
\end{figure}

The increase of HBT radii with multiplicity has also been observed previously in $p-\bar{p}$ collisions by the E735
Collaboration~\cite{Alexopoulos:1992iv}.  While in $A+A$ collisions, this is naturally related to increasing length
scales in the entrance channel geometry, Pai\'{c} and Skowro\'{n}ski~\cite{Paic:2005cx} postulate that jet dynamics, rather than bulk
properties, drive this dependence in the $p+\bar{p}$ system.  Within a simple model of hadronization, they can
reproduce the E735 multiplicity dependence, and make predictions of similar multiplicity dependence for $p+p$ collisions
at the top LHC energy.  However, since the expectation of increasing length scales with multiplicity
seems to be rather generic to {\it all} scenarios, it will be interesting to see these predictions
expanded to more differential measures-- say, the multiplicity {\it and} $p_T$ dependence, probing both
aspects of Equation~\ref{eq:Factorization}.  This should allow a more discriminating comparison between
models, allowing some to be ruled out.

Such differential predictions have very recently been performed by Humanic~\cite{Humanic:2006ib}, in the context of a Pythia+hadronic~rescattering
(through HRM) scenario.  It is found that, contrary to some expectations, hadronic rescattering is crucial to understand the
$M \otimes p_T$ dependence of the E735 data.  Reproducing the data requires the assumption of a surprisingly
short hadronization timescale ($\sim 0.1$~fm); longer timescales do not allow sufficient hadronic rescattering needed
to describe the $p_T$-dependence.  The prediction of the model for the highest multiplicity $p+p$ collisions at $\sqrt{s}=1.4$~TeV are shown in
Figure~\ref{fig:PythiaHRMpp}.  
%%%%%%Clearly, the predictions are very sensitive to the hadronization time.

\section{Summary}
\label{sec:summary}

From an extended and deformed initial state, to medium-induced energy loss of partons, to collective flow, to
deconfinement itself,
space-time considerations underlie any understanding of the bulk system created in a heavy ion collision.
Transport model calculations that simulate the evolution of this highly dynamic system, produce space-time
configurations which reflect basic physical quantities like the Equation of State, viscosity, and degree
of equilibration.

The most direct way to get at this space-time information is through femtoscopic measurements.
With the availability of first nuclear beams a quarter century ago, a femtoscopic measurement was 
considered a success if it could demonstrate that the overall scale of the system was on the order of a few fermi.
Since then, much improved experimental techniques and more precise data, coupled with advances in theory
and phenomenology, have made femtoscopy a precision tool.  A comprehensive program has resulted in multi-dimensional
(e.g. ``out-side-long'') pion correlation functions as a function of $y$, $p_T$ and $\phi$; similar measurements
for other particle species combinations have been steadily accumulating.

These systematic measurements allow the extraction of the size, shape, orientation, timescales, and even
(for non-identical particle correlations) the positions of the homogeneity regions.
The momentum- and species-dependences of these quantities reveal the nature and magnitude of collective
behavior through the signature geometric substructure it generates.

Generically, the data indicate an emitting source whose overall space and time scales are determined
by the final-state chemistry.  A long-range component of the emission function has been identified
and is consistent with expections from contributions from resonance decay.
%%; at the highest energies pions dominate, and the event multiplicity is
%%what is important.  
The strong space-momentum correlations probe longitudinal, radial, and elliptical
collective flow.  They reveal a system that is approximately boost-invariant near midrapidity, 
dominated by a position-dependent collective boost field.  For non-central collisions, the
system's shape evolves away from its initial out-of-plane extended geometry, due to anisotropic
flow fields coupled with finite evolution time.
This flow field is {\it the} main feature reflecting the response of the bulk system to the extreme conditions of the collision.
The flow-dominated scenario is quantitatively
consistent with momentum-only measurements like $p_T$ spectra, rapidity distributions, and emission
anisotropy (e.g. $v_2$).  However, any inference of the structure of the flow field, based on
the momentum-only data alone, would be a faith-based exercise.

Moving beyond the generic scenario requires comparison with transport calculations.
After initial consternation followed by detailed study, it now appears that the femtoscopic data 
set for heavy ion collisions from RHIC is on the verge of being satisfactorily explained. The source 
of much of the mystery in understanding source sizes at RHIC derived from a conspiracy of model features, 
which had been neglected or poorly implemented in previous dynamic treatments. Although none of these 
were individually sufficient to explain the HBT puzzle, the conspiracy of several corrections to push 
predictions in the same direction appears to bring predictions and data into reasonable agreement. 
Morale and optimism are high, but significant work remains to be done as the success has only been 
demonstrated for $\pi\pi$ correlations. Simultaneous analyses of azimuthally sensitve HBT
and of non-identical particle correlations are needed and underway. Furthermore, while toy parameterizations
can simultaneously reproduce essentially all soft-sector data, it has yet 
to be demonstrated that femtoscopic observables can be explained with the same model as elliptic flow. 
Only then, can victory be declared.  We are now, however, close.

Transport models will be tested even more strenuously, as heavy ion data at the LHC becomes available.
Several groups have generated femtoscopic predictions from
Boltzmann and hydrodynamic calculations; thus far,
they all pertain to two-pion interferometry only.  Compared to those at RHIC, geometric scales are
expected to increase on the order of 35\%, though there are significant differences between the
model predictions.  All predict a break with the ``universal'' $k_T$-dependence observed at
RHIC and lower energies.  In particular, they predict steeper fall-off of the transverse radii,
due to much larger predicted transverse flow.  This increased flow is further reflected in predictions
of less-steep single particle spectra.

The confidence with which we now state our understanding of heavy ion collisions may be threatened
by similar data for small sources. Most of the same features and trends observed in heavy ion collisions, 
and often attributed to collective phenomena, persist for the smallest $p+p$ source sizes. An analogous
situation may also exist for other observables: there are suggestions that, modulo
finite number effects, chemical yields and species-dependent $p_T$ spectra from hadronic collisions
are quite similar to those from heavy ion reactions.
This challenges our understanding at both the quantitative and qualitative level, and might signal 
the need for a new theoretical paradigm for dynamics and hadronization of small systems.
Certainly, it raises issues for further research at both RHIC and the LHC, where apples:apples comparisons
are possible.

%%%%%\bibliographystyle{annrev}
%%%%%\bibliography{citations}

\end{document}